\documentclass[prx,twocolumn,amsmath,amssymb,superscriptaddress]{revtex4}

\usepackage[pdftex]{graphics}
\usepackage{graphicx}
\usepackage{times}

\newcommand{\nair}{Na$_2$IrO$_3$}
\newcommand{\liir}{$\alpha$-Li$_2$IrO$_3$}
\newcommand{\rucl}{$\alpha$-RuCl$_3$}

\begin{document}

\title{Fingerprints of Kitaev physics in the magnetic excitations of honeycomb iridates}

\author{A. Revelli}
\affiliation{II. Physikalisches Institut, Universit\"{a}t zu K\"{o}ln, Z\"{u}lpicher Strasse 77, D-50937 K\"{o}ln, Germany}
\author{M. Moretti Sala}
\affiliation{European Synchrotron Radiation Facility, BP 220, F-38043 Grenoble Cedex, France}
\affiliation{Dipartimento di Fisica, Politecnico di Milano, Piazza Leonardo da Vinci 32, I-20133 Milano, Italy}
\author{G. Monaco}
\affiliation{Dipartimento di Fisica, Universit\`{a} di Trento, via Sommarive 14, 38123 Povo (TN), Italy}
\author{C. Hickey}
\affiliation{Institut f{\"u}r Theoretische Physik, Universit\"{a}t zu K\"{o}ln, Z\"{u}lpicher Strasse 77, D-50937 K\"{o}ln, Germany}
\author{P. Becker}
\affiliation{Abteilung Kristallographie, Institut f\"{u}r Geologie und Mineralogie, Z\"{u}lpicher Strasse 49b, D-50674 K\"{o}ln, Germany}
\author{F. Freund}
\author{A. Jesche}
\author{P.~Gegenwart}
\affiliation{Experimental Physics VI, Center for Electronic Correlations and Magnetism, University of Augsburg, 86159 Augsburg, Germany}
\author{T.~Eschmann}
\author{F. L. Buessen}
\author{S. Trebst}
\affiliation{Institut f{\"u}r Theoretische Physik, Universit\"{a}t zu K\"{o}ln, Z\"{u}lpicher Strasse 77, D-50937 K\"{o}ln, Germany}
\author{P. H. M. van Loosdrecht}
\affiliation{II. Physikalisches Institut, Universit\"{a}t zu K\"{o}ln, Z\"{u}lpicher Strasse 77, D-50937 K\"{o}ln, Germany}
\author{J. van den Brink}
\affiliation{Institute for Theoretical Solid State Physics, IFW Dresden, Helmholtzstrasse 20, 01069 Dresden, Germany}
\author{M. Gr\"{u}ninger}
\affiliation{II. Physikalisches Institut, Universit\"{a}t zu K\"{o}ln, Z\"{u}lpicher Strasse 77, D-50937 K\"{o}ln, Germany}

\date{May 31, 2019}

\begin{abstract}
In the quest for realizations of quantum spin liquids, the exploration of Kitaev materials -- spin-orbit entangled 
Mott insulators with strong bond-directional exchanges -- has taken center stage. However, in these materials the 
local spin-orbital $j$\,=\,$1/2$ moments typically show long-range magnetic order at low temperature, 
thus defying the formation of a spin-liquid ground state. 
Using resonant inelastic x-ray scattering (RIXS), we here report on a proximate spin liquid regime
with clear fingerprints of Kitaev physics in the magnetic excitations of the honeycomb iridates \liir{} and \nair.
We observe a broad continuum of magnetic excitations 
that persists up to at least 300\,K, more than an order of magnitude larger than the magnetic ordering temperatures. 
We prove the magnetic character of this continuum by an analysis of the resonance behavior. 
RIXS measurements of the dynamical structure factor for energies within the continuum show that 
dynamical spin-spin correlations are restricted to nearest neighbors.
Notably, these spectroscopic observations are also present in the magnetically ordered state for excitation energies
above the conventional magnon excitations.
Phenomenologically, our data agree with inelastic neutron scattering results on the related honeycomb 
compound \rucl, establishing a common ground for a proximate Kitaev spin-liquid regime in these materials. 
\end{abstract}

\maketitle

\section{Introduction}

Spin liquids are elusive states of quantum matter in which quantum fluctuations prevent the formation of 
long-range magnetic order even at $T$\,=\,0 \cite{Broholm19}. Theory distinguishes various types of 
spin liquids, e.g., gapless and gapped variants with topological order \cite{Savary17,Knolle19}. 
A particularly interesting version arises from exchange frustration as described by the Kitaev model 
on tricoordinated lattices \cite{Kitaev06,OBrien16}, which exhibits fractional excitations in the form of
itinerant Majorana fermions and static visons \cite{Hermanns18}. 
\begin{figure}[b!]
	\centering
	\includegraphics[width=1\columnwidth]{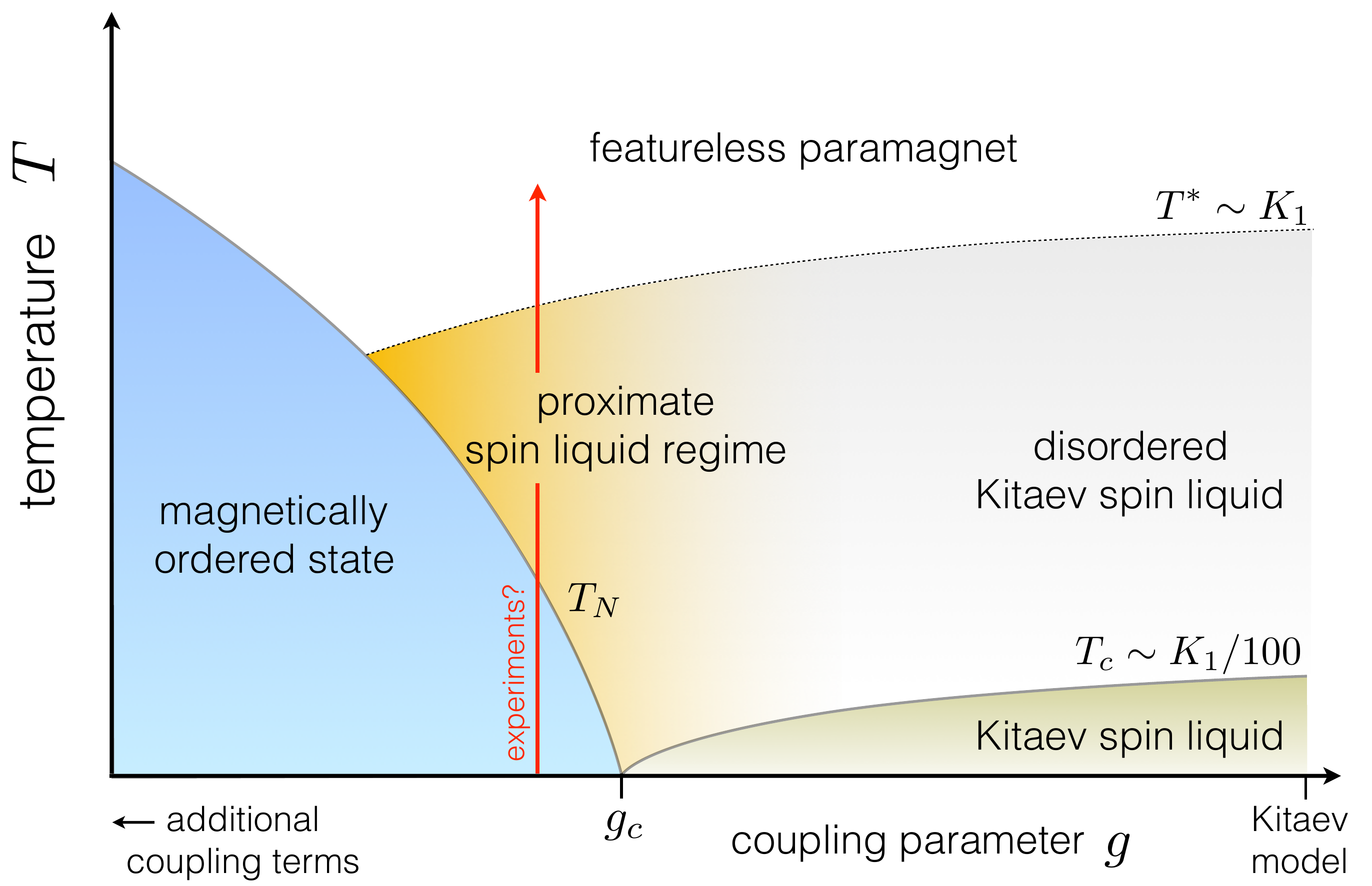}
	\caption{\textbf{Proximate spin liquid regime} in a hypothesized phase diagram 
	of the Kitaev model in the presence of additional magnetic interactions.
	In the pure Kitaev model (right), the ordered $Z_2$ quantum spin-liquid ground state is separated 
	from the featureless, high-temperature paramagnet by a disordered $Z_2$ spin liquid. 
	Long-range magnetic order is expected for sufficiently large strength of additional exchange terms (left).
	Above the magnetic ordering transition, an intermediate temperature regime might persist, 
	in which the system resides in a ``proximate spin liquid regime", where it exhibits fingerprints of 
	Kitaev physics up to temperatures of the order of the Kitaev coupling.
}
\label{fig_sketch_phase}
\end{figure}
Realizations of such Kitaev spin liquids are sought in Mott insulators with spin-orbit-entangled 
$j$\,=\,1/2 moments \cite{Witczak14} such as the honeycomb iridate H$_3$LiIr$_2$O$_6$  \cite{Kitagawa18} 
which does not show magnetic order down to at least 50\,mK \cite{FootnoteDisorder}.
The related compounds \liir, \nair, and \rucl{} all show dominant bond-directional Kitaev 
interactions \cite{Chun15,Rau2016,Winter17rev} but order magnetically below 7\,-15\,K,
which is commonly attributed to additional coupling terms, e.g., Heisenberg and off-diagonal 
interactions \cite{Rau2016,Winter17rev,Trebst2017}.

This raises the question whether these materials are, by virtue of their dominant bond-directional interactions,
destined to form Kitaev spin liquids, which however is impeded at low temperatures by spurious 
additional couplings.
This situation can be compared to quasi-1D quantum spin systems such as KCuF$_3$ 
showing 3D long-range order at low temperature due to a residual inter-chain coupling. 
Despite magnetic ordering, the excitation spectrum can, however, still be described by a continuum 
of fractional spinons for energies larger than the energy scale of the inter-chain coupling \cite{Lake13}, 
with the low-energy magnons taking only a small fraction of the total spectral 
weight. At high energy, the spinon continuum is not affected by the ordering temperature $T_N$ and persists 
to temperatures far above $T_N$.
For the Kitaev materials at hand, this suggests the existence of a {\em proximate spin-liquid regime} 
above the long-range ordering temperature in which the physics of fractionalized excitations manifests itself, 
see Fig.\ \ref{fig_sketch_phase}. 
Indeed, \rucl{} has been described as a proximate spin liquid based on inelastic neutron scattering data 
\cite{Banerjee16,Banerjee17} which revealed a continuum of magnetic excitations, smeared out both in energy and momentum space, 
which persists far above the N\'eel temperature $T_N$. 
The large width in energy is attributed to the fractional character of excitations; 
the sinusoidal momentum dependence of the dynamical structure factor reflects the dominance 
of nearest-neighbor correlations. Together, these count as clear fingerprints of spin-liquid behavior 
as reported in inelastic neutron scattering on the Kagome lattice antiferromagnet Herbertsmithite \cite{Han12} 
and Ca$_{10}$Cr$_7$O$_{28}$ \cite{Balz16}. 
The proximity of \rucl{} to a Kitaev spin liquid was most prominently highlighted by the 
observation of a half-integer quantized thermal Hall effect attributed to fractionalized Majorana fermions 
in an external magnetic field suppressing long-range order \cite{Kasahara18}.  
Also a low-energy Raman continuum was attributed to Majorana fermions \cite{Nasu16,Sandilands15}.

Here, we address the question whether the existence of a proximate spin liquid regime is a generic
feature of Kitaev materials by exploring the $j$\,=\,1/2 honeycomb iridates \nair{} and \liir{}.
According to \textit{ab-initio} calculations \cite{Winter17rev}, the iridates and in particular \nair{} 
are expected to be closer to the pure Kitaev limit than \rucl. However, experimental studies of the 
magnetic excitations in the honeycomb iridates are challenging and thus scarce. 
Powder inelastic neutron scattering data of \nair{} and \liir{} show magnon modes at 
a few meV \cite{Choi12,Choi19}, but further investigations are obstructed by the strong neutron absorption 
of Ir. Resonant inelastic x-ray scattering (RIXS) has been used to study the magnetic excitations of 
corner-sharing iridates \cite{Kim12,Gretarsson16,Pincini17,MorettiSr3,Lu17}  
but the much smaller energy scale of edge-sharing honeycomb iridates is comparable 
to the state-of-the-art energy resolution of 24\,meV \cite{Moretti13,Moretti18}. 
In \nair, a broad inelastic RIXS feature peaking at about 30 to 50\,meV 
was previously attributed to the sum of magnons and phonons \cite{Gretarsson13PRB}.

To explore the existence of a proximate spin liquid regime in these honeycomb iridates
we probe their magnetic excitations using RIXS.
We resolve the character of the low-lying excitations by comparing \nair{} and \liir{} 
as well as by studying the resonance behavior and the $\mathbf{q}$ dependence of the 
RIXS intensity. We find clear fingerprints of proximate spin-liquid behavior:
(i) a broad continuum of quasi-2D magnetic excitations centered around the $\Gamma$ point 
of the first Brillouin zone, 
(ii) the insensitivity of this continuum to the magnetic ordering temperature $T_N$ 
and the persistence of the continuum up to temperatures as high as 300\,K, i.e., far above $T_N$, 
and 
(iii) dynamical spin-spin correlations restricted to nearest-neighbor bonds. 
The latter is a key feature of Kitaev exchange on tricoordinated lattices. 
Observations (i)-(iii) are comparable to the neutron results for \rucl{} \cite{Banerjee17} discussed above, 
establishing a common experimental ground for these $j$\,=\,1/2 honeycomb compounds. 
Going beyond the thermodynamic perspective of the phase diagram sketched in Fig.\ \ref{fig_sketch_phase}, 
the magnetic continuum for excitation energies $\hbar \omega \gtrsim k_B T_N$ shows fingerprints 
of proximate Kitaev spin-liquid behavior even below the magnetic ordering temperature 
-- a {\em spectroscopic} proximate spin liquid regime.

\section{Crystal synthesis and RIXS measurements}

Single crystals of \nair{} can be grown by solid-state synthesis out of a polycrystalline bed \cite{Singh10}.
Single crystals of several mm size were grown from Na$_2$CO$_3$ and Ir in a ratio of 1.5/1 at 1423 K.\@ 
After cooling to room temperature the crystals were separated mechanically from the remaining polycrystalline 
sintered compact.
Sub-millimeter single crystals of \liir{} can be grown from separated educts \cite{Freund16}. 
\nair{} and \liir{} both crystallize in the monoclinic $C2/m$ space group \cite{Choi12}, 
display opposite magnetic anisotropy, and order antiferromagnetically in zigzag and spiral states, 
respectively, near 15\,K \cite{Winter17rev}. 
Crystal quality for the present study was checked by XRD and magnetic measurements.
We use the reciprocal lattice of the monoclinic setting. In this framework, $\Gamma$ points of the 
2D honeycomb net are given by, e.g., (0 0), \mbox{(1 1)}, or (2 0). 

\begin{figure}[t!]
	\centering
	\includegraphics[width=1\columnwidth]{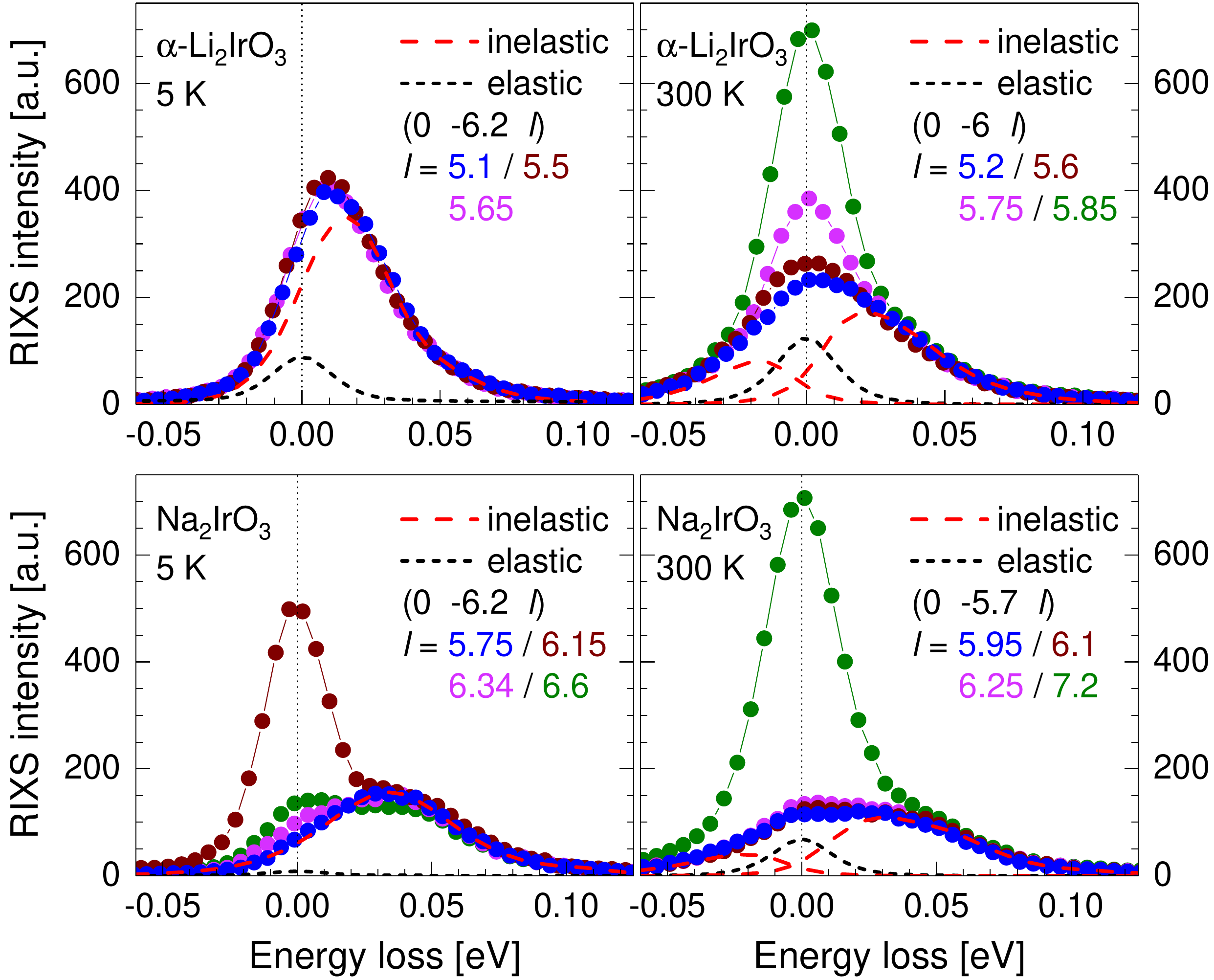}
	\caption{\textbf{Disentangling inelastic features from the elastic line.}
		Low-energy RIXS spectra of \liir{} (top) and \nair{} (bottom) 
		at 5\,K (left) and 300\,K (right) in the vicinity of $\mathbf{q}$\,=\,(0 -6 $l$).  
		The width of the elastic line is $\Delta E$\,=\,24\,meV.\@ 
		Its intensity increases close to a Bragg peak and additionally depends on the 
		scattering angle $2\theta$. In particular, the elastic line is strongly suppressed for $2\theta$\,=\,$90^\circ$. For the selected values of $l$, $2\theta$ covers the range from 
		$90^\circ$ to $110^\circ$. 
		The data above about 20\,meV are nearly independent of $l$, demonstrating the presence of
		inelastic features with a quasi-2D character. 
	}
\label{fig_all_raw}
\end{figure}

RIXS was measured at the Ir $L_3$ edge at beamline ID20 at the ESRF.\@ 
The measurements were performed with incident $\pi$ polarization in the horizontal scattering plane 
in order to suppress the elastic line for $2\theta$ close to 90$^\circ$. 
The energy resolution equals $\Delta E$\,=\,24\,meV, 
while the resolution of the transferred momentum $\mathbf{q}$ amounts to about 1/5 of the Brillouin zone. 
The incident energy was tuned to the Ir $2p_{3/2}\! \rightarrow \! 5d_{t_{2g}}$ absorption edge, 
i.e., to resonate with the $t_{2g}$ manifold forming the $j$\,=\,$1/2$ state. 
For both compounds, the measurements were performed on a (001) surface. 
The samples were oriented such that the $b$ axis was in the scattering plane 
and the $a$ axis perpendicular to it.

\section{Experimental results}

\subsection{Broad continuum of excitations}

One general challenge for RIXS studies of magnetic excitations in  the honeycomb iridates 
is that the relevant excitation energies one has to cope with are comparably low. 
Therefore, we first address how, in our RIXS measurements, we distinguish inelastic features from the elastic line. 
Figure \ref{fig_all_raw} shows exemplary low-energy RIXS spectra for \liir{} (top) 
and \nair{} (bottom) at 5\,K (left) and 300\,K (right). The elastic line at zero energy loss 
with a width of $\Delta E$\,=\,24\,meV overlaps with inelastic features that extend up to about 100\,meV.\@ 
Below we will show that this inelastic contribution corresponds to magnetic excitations. 
To disentangle inelastic features and the elastic line, we change the intensity of the latter by varying 
the scattering angle $2\theta$ in the range close to $90^\circ$. This is achieved by varying $l$ in the 
transferred momentum $\mathbf{q}$\,=\,\mbox{($h$ $k$ $l$)}. 
Ideally, the elastic line is fully suppressed for $2\theta$\,=\,$90^\circ$. 
We fit the spectra for different $l$ by a sum of elastic line (black dashed) and inelastic features 
(red dashed) using the Pearson VII line shape \cite{Wang05}. 
At 300\,K, the fit includes an Anti-Stokes contribution at negative energy loss, 
where the Anti-Stokes-to-Stokes intensity ratio is given by $\exp(-\hbar\omega/k_B T)$.
In the fit, only the intensity of the elastic line is allowed to vary with $l$. 
The insensitivity to $l$ of the inelastic features reflects the quasi-2D character of the excitations.

\begin{figure}[t]
	\centering
	\includegraphics[width=0.95\columnwidth]{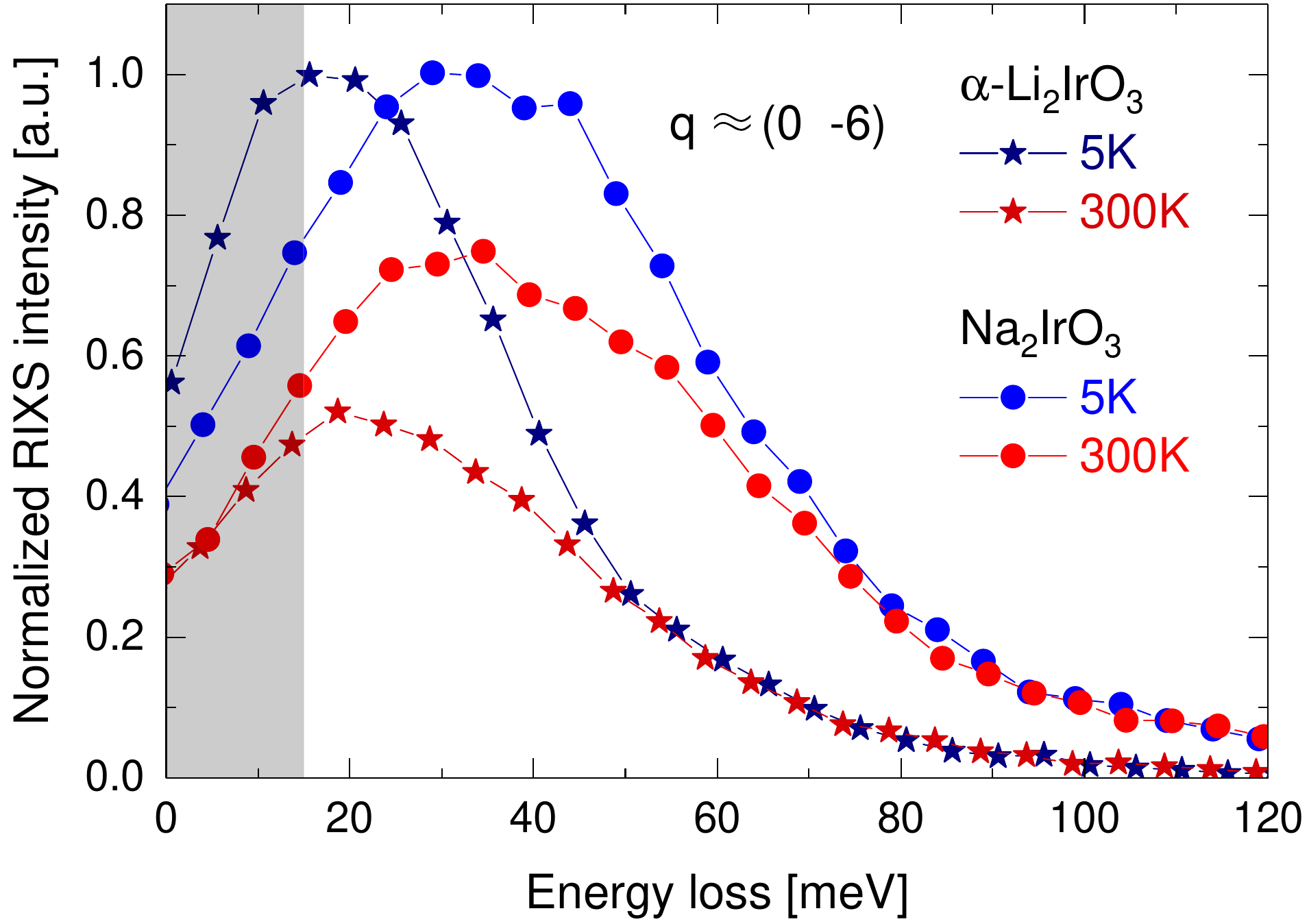}
	\caption{\textbf{Stokes contribution of RIXS spectra in the vicinity of ($h$ $k$)\,=\,(0 -6).} 
		The data depict the result of the analysis shown in Fig.\ \ref{fig_all_raw} 
		and were normalized to the maximum value observed at 5\,K. 
		We find a broad continuum which persists up to high temperature. 
		Below 15\,meV (gray), the uncertainty increases since the precise intensity of the elastic line 
		can only be estimated. }
	\label{fig_all_Stokes}
\end{figure}

The resulting Stokes contribution for transferred in-plane momenta close to (0 -6) is depicted in 
Fig.\ \ref{fig_all_Stokes}. Note that (0 -6) is equivalent to a $\Gamma$ point of the reciprocal lattice 
of the honeycomb net. At 5\,K, we find a broad excitation continuum peaking at about 15\,meV in \liir{} 
and 35\,meV in \nair. 
A continuum character is suggested by the large width of more than 50\,meV, which is much larger than 
the experimental resolution $\Delta E$\,=\,24\,meV, and by the featureless line shape. 
A continuum reflects a two-particle or multi-particle character of the excitation, 
a hallmark of fractional excitations.

\subsection{Temperature dependence}

\begin{figure}[t]
	\centering
	\includegraphics[width=0.95\columnwidth]{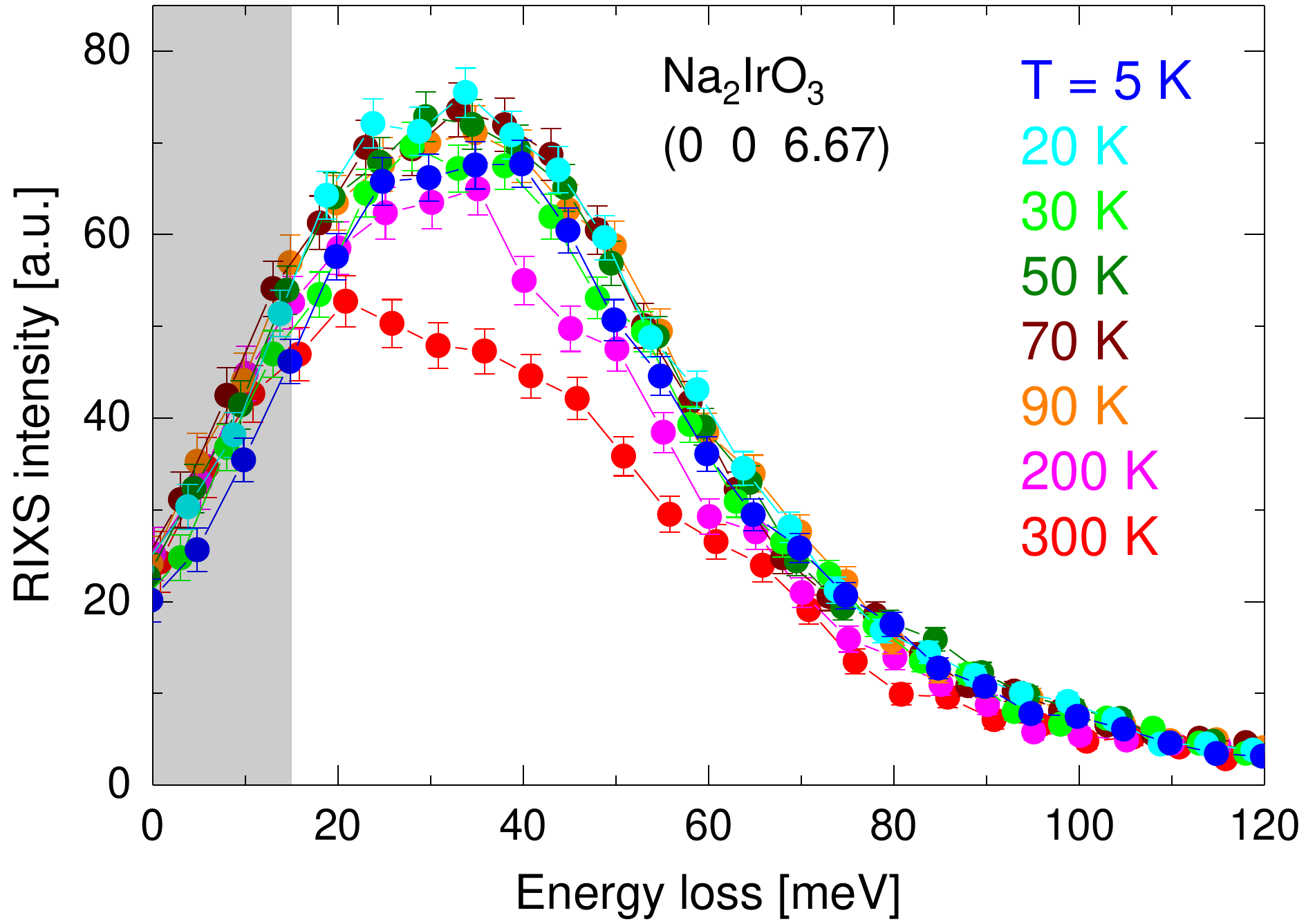}
	\caption{\textbf{Temperature dependence of the continuum in \nair.} 
		The plot depicts the Stokes contribution at (0 0 6.67).
		Data were normalized by the integrated RIXS intensity from 0.3 to 1.1\,eV.\@ 
		The continuum hardly changes between 5\,K and 90\,K and persists up to 300\,K.\@ 
		In particular, the continuum is insensitive to the  N\'eel temperature $T_N$\,=\,15\,K.\@
		Below 15\,meV (gray), the uncertainty increases, cf.\ Fig.\ \ref{fig_all_Stokes}. 
	} 
	\label{fig_Na_T}
\end{figure}
 
At 300\,K, the RIXS data are stunningly similar to the 5\,K result, the main change being a reduction 
of the peak intensity, see Fig.\ \ref{fig_all_Stokes}. 
The temperature-driven reduction is larger in \liir, in agreement with the smaller energy scale. 
For \nair, the striking robustness with respect to temperature is emphasized in Fig.\ \ref{fig_Na_T} 
which shows that the RIXS peak for $q$\,=\,(0 0 6.67) hardly changes between 5\,K and 90\,K\,$\approx 6T_N$.  

The insensitivity to temperature across $T_N$\,=\,15\,K and the robustness up to 6\,$T_N$
shows that the RIXS continuum cannot be explained in terms of magnons of the magnetically 
long-range ordered state observed below $T_N$. Also the quasi-2D character speaks against magnons 
of the 3D ordered state. As discussed in the next section, 
our RIXS data nevertheless provide strong evidence for a {\em magnetic} character of the continuum, 
pointing towards a more intriguing origin. A dominant Kitaev exchange is well accepted for the 
honeycomb iridates \cite{Chun15,Rau2016,Winter17rev}, 
in which long-range order is thought to arise due to additional weaker 
coupling terms beyond the pure Kitaev model \cite{Rau2016,Winter17rev,Trebst2017}.
This suggests an analogy with quasi-one dimensional systems 
which show long-range order at low temperature due to finite inter-chain couplings. Nevertheless the 
excitations at higher energy are well described as fractional spinon excitations, 
\textit{both below and above} the ordering temperature.
For the iridates, this suggests that the proximate Kitaev spin liquid leaves its 
fingerprints not only in the magnetic excitations above the ordering temperature, as sketched in 
Fig.\ \ref{fig_sketch_phase}, but also in the high-energy magnetic excitations for temperatures below 
the long-range ordering temperature. In this scenario, long-range order mainly affects 
the excitations at lower energy, i.e., below 20\,meV in \nair. 
The high-energy Kitaev-related excitations are much more robust against temperature than conventional magnons.

\subsection{Magnetic character and resonance behavior }

\begin{figure}[t]
	\centering
	\includegraphics[width=0.95\columnwidth]{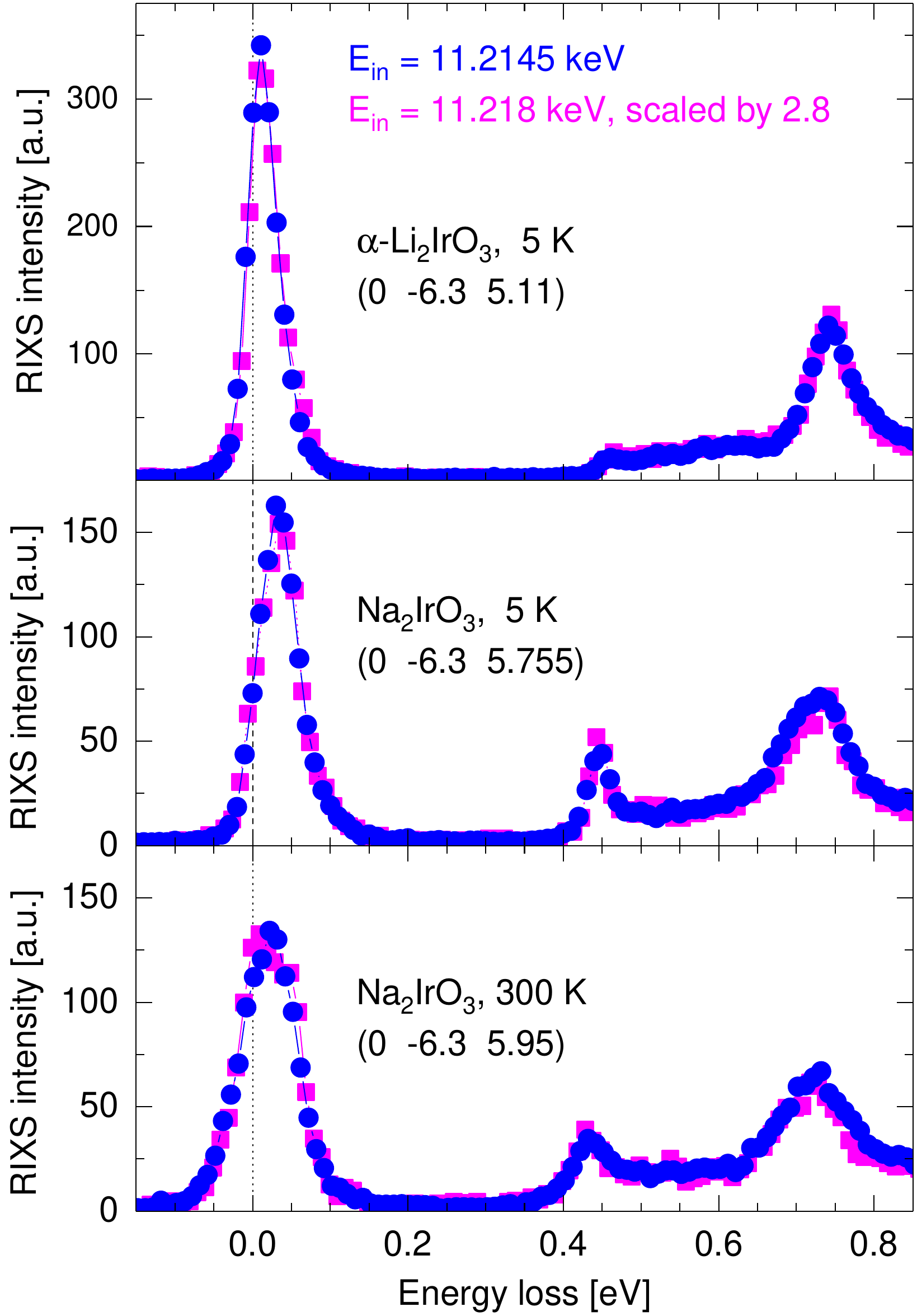}
	\caption{\textbf{Resonance behavior of RIXS spectra, establishing the magnetic character of the 
			low-energy continuum.}
		RIXS was measured on \liir{} and \nair{} for incident energies of 11.2145\,keV and 11.218\,keV, equivalent 
		to constant-$E_{\rm in}$ cuts through the resonance maps shown in Fig.\ \ref{fig_resonance_maps}. 
		To suppress the elastic line, the transferred momentum was chosen such that 
		$2\theta$\,$\approx$\,$90^\circ$.		
		Maximum resonance enhancement is obtained at the $t_{2g}$ resonance with $E_{\rm in}$\,=\,11.2145\,keV, 
		see Fig.\ \ref{fig_resonance_cuts}. 
		Data for $E_{\rm in}$\,=\,11.218\,keV were scaled up by a factor 2.8 to compensate for detuning 
		$E_{\rm in}$ away from the $t_{2g}$ resonance. 
		After scaling, the spectra fall on top of each other over the full measurement range, 
		demonstrating that the excitation occurs via a direct RIXS process for all observed features.  
	} 
	\label{fig_resonance}
\end{figure}

At first sight, a magnetic interpretation of excitations that show such low sensitivity to temperature 
is counterintuitive and requires a solid experimental foundation, which we will give in the following. 
Our RIXS spectra of \nair{} shown in Fig.\ \ref{fig_Na_T} agree very well with the data reported by 
Gretarsson \textit{et al.} \cite{Gretarsson13PRB}. They interpreted their data in terms of a series 
of overlapping phonons at high temperature and as a sum of phonons and magnons at 
low temperature \cite{Gretarsson13PRB}. 
In their scenario, the small temperature dependence of the intensity on the Stokes side is due to 
an accidental balance of the phonon intensity increasing and the magnon intensity decreasing with 
increasing temperature. 
A first argument against this scenario is provided by the comparison of the data for isostructural 
\liir{} and \nair, see Fig.\ \ref{fig_all_Stokes}.
Phonons are expected to be very similar in these 
two compounds, with an enhanced energy scale in \liir{} due to the smaller mass of Li compared to Na, 
as observed in optical spectroscopy \cite{Hermann17}. 
In contrast, the energy scale is markedly lower in \liir{} in the RIXS data, 
questioning the phonon interpretation.

\begin{figure}[t]
	\centering
\includegraphics[width=0.508\columnwidth]{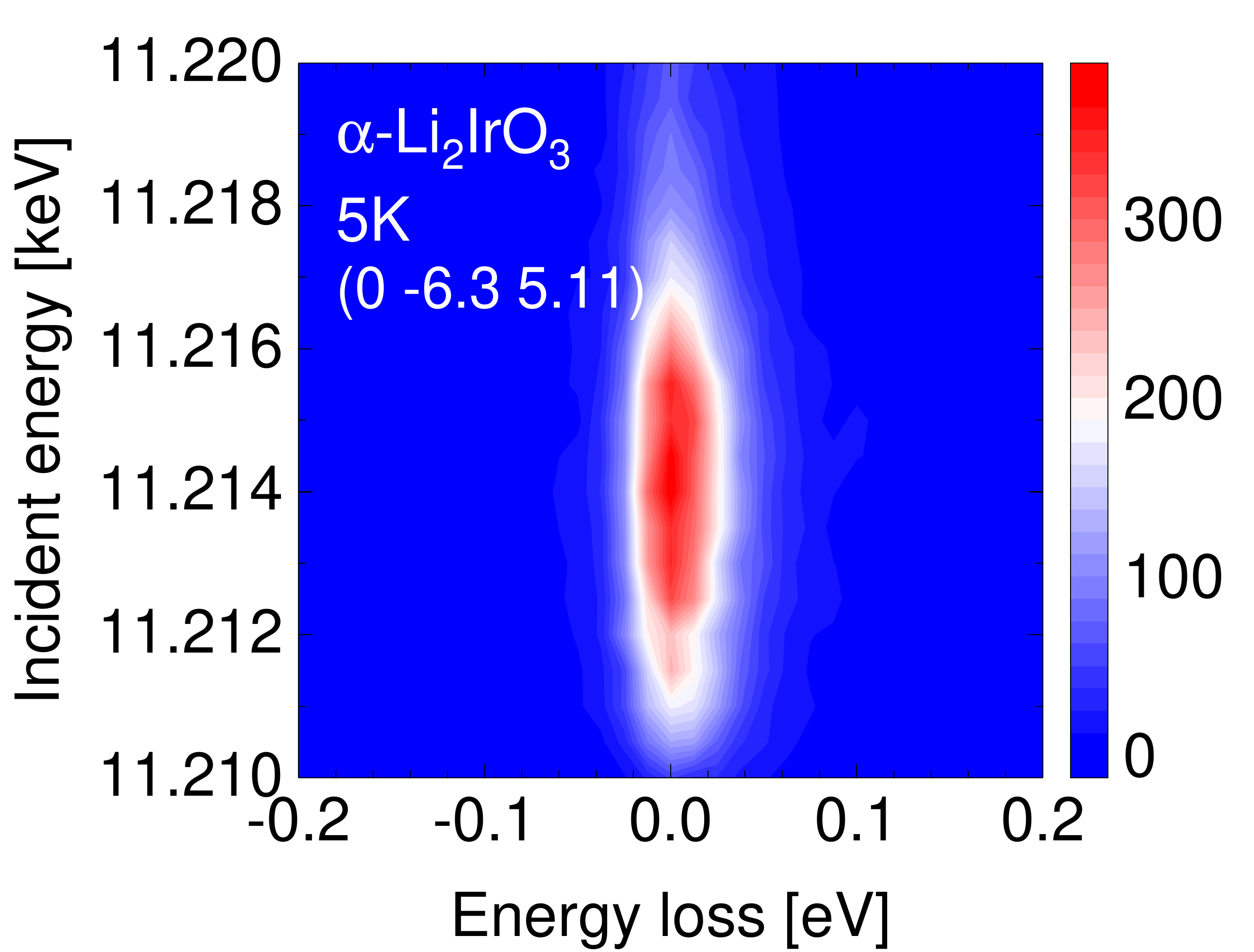}
\includegraphics[width=0.472\columnwidth]{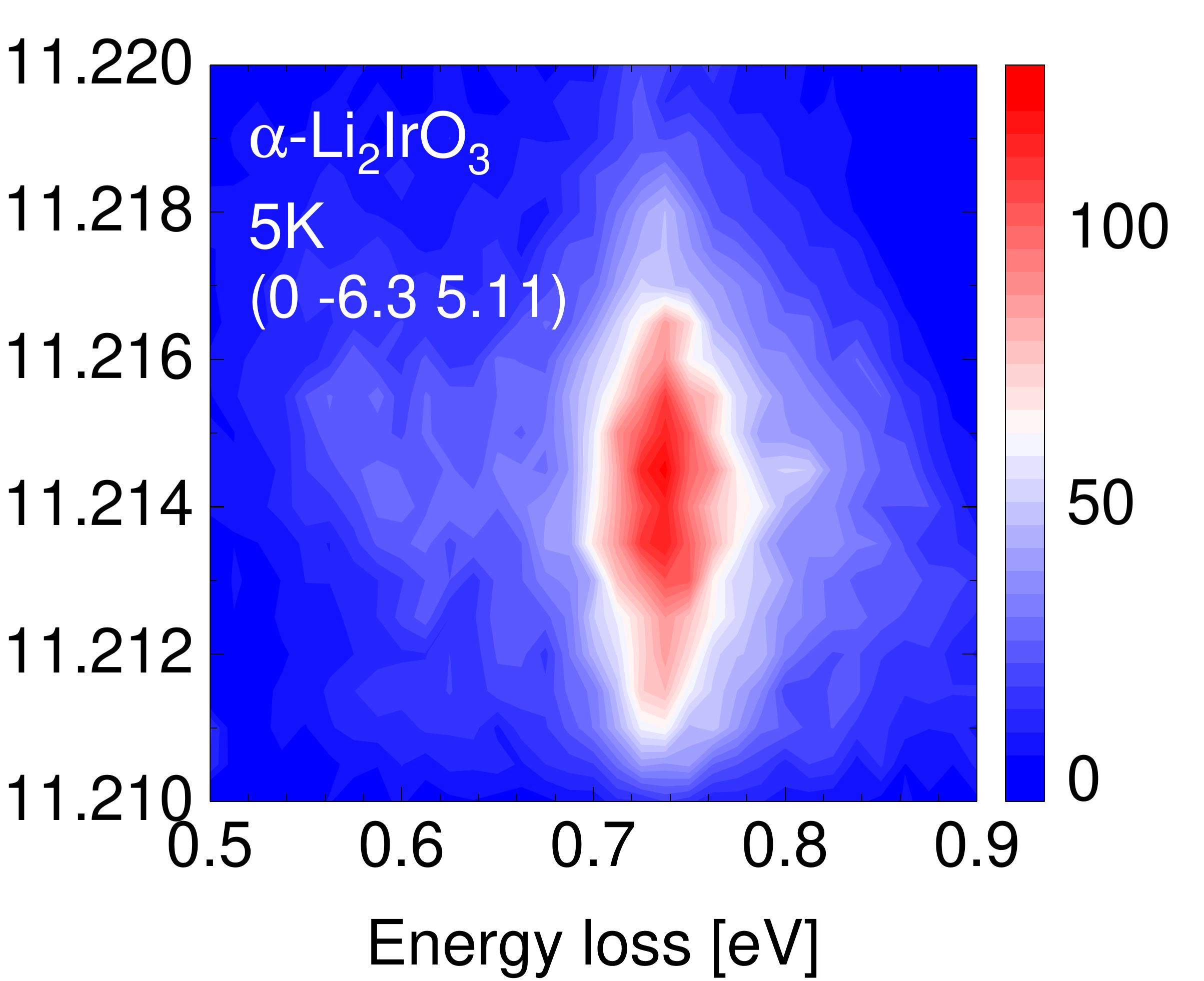}
\includegraphics[width=0.508\columnwidth]{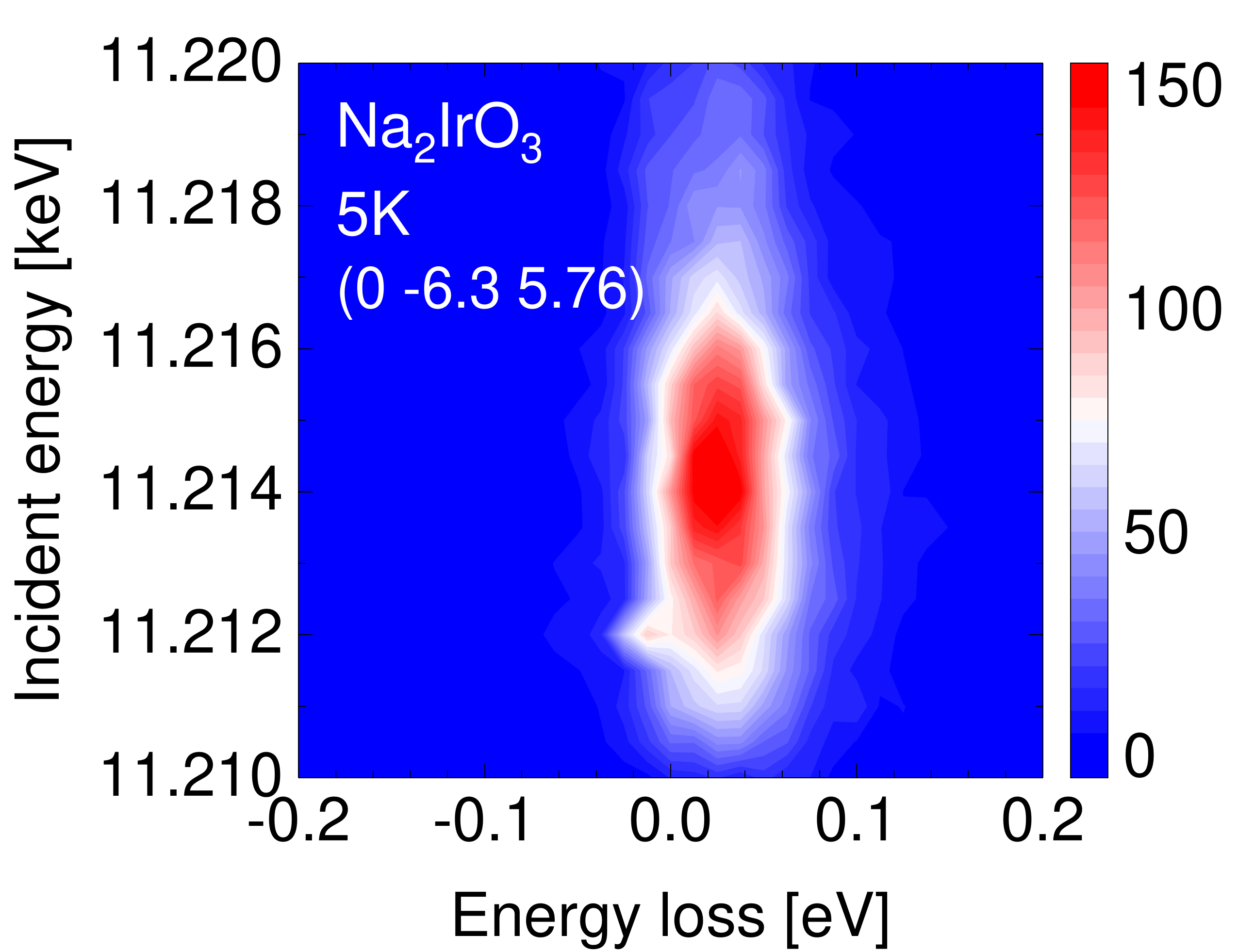}
\includegraphics[width=0.472\columnwidth]{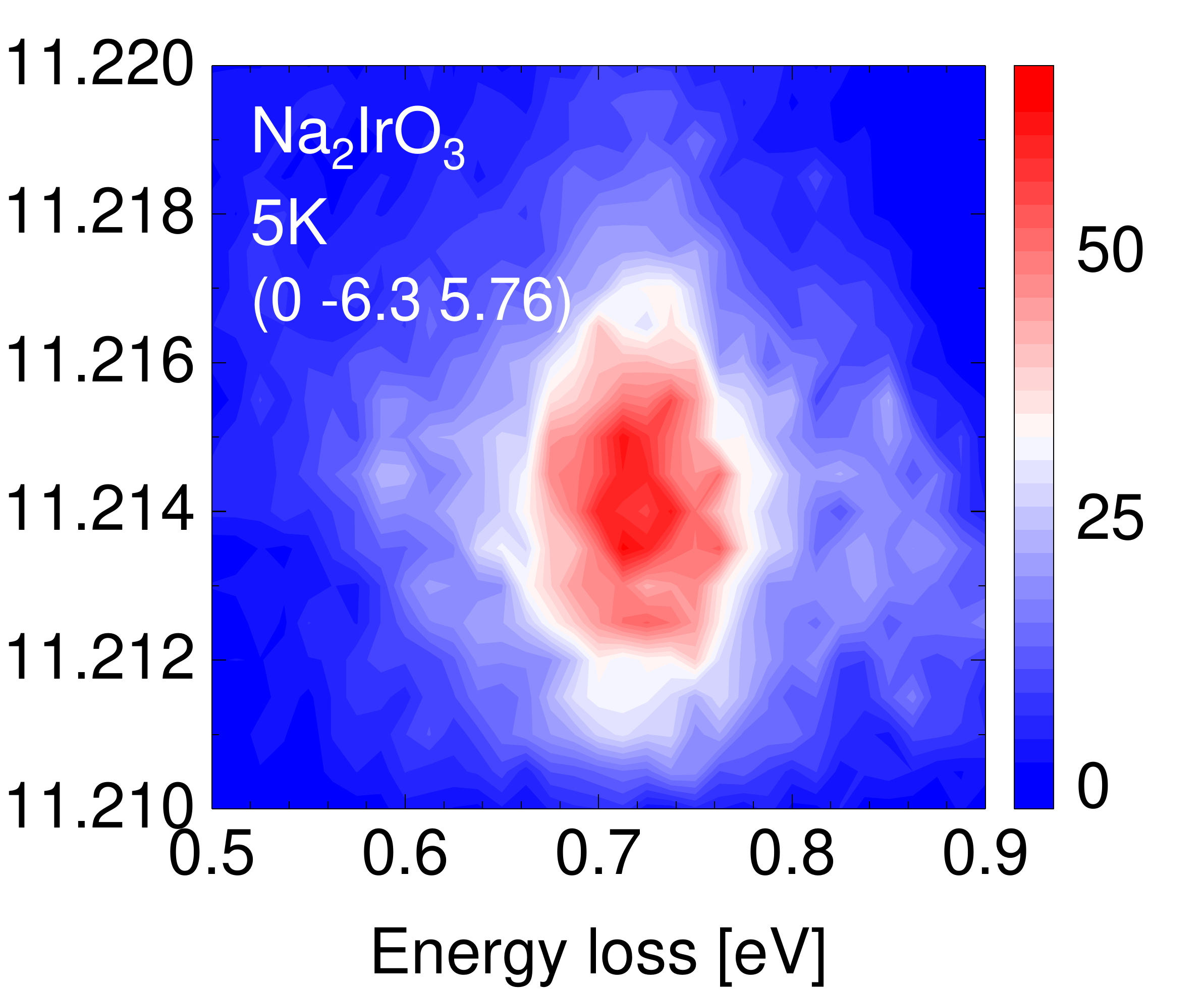}
\includegraphics[width=0.508\columnwidth]{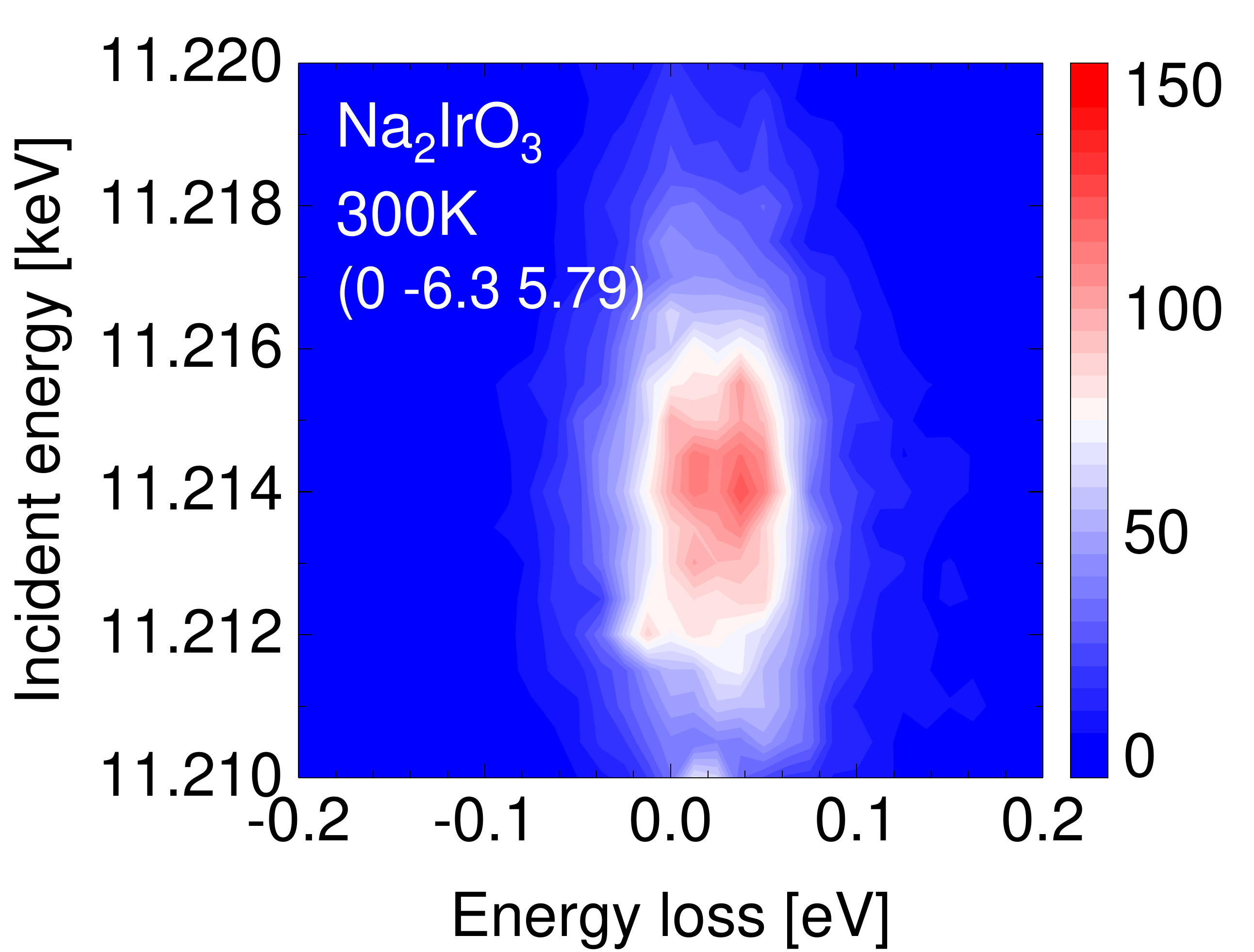}
\includegraphics[width=0.472\columnwidth]{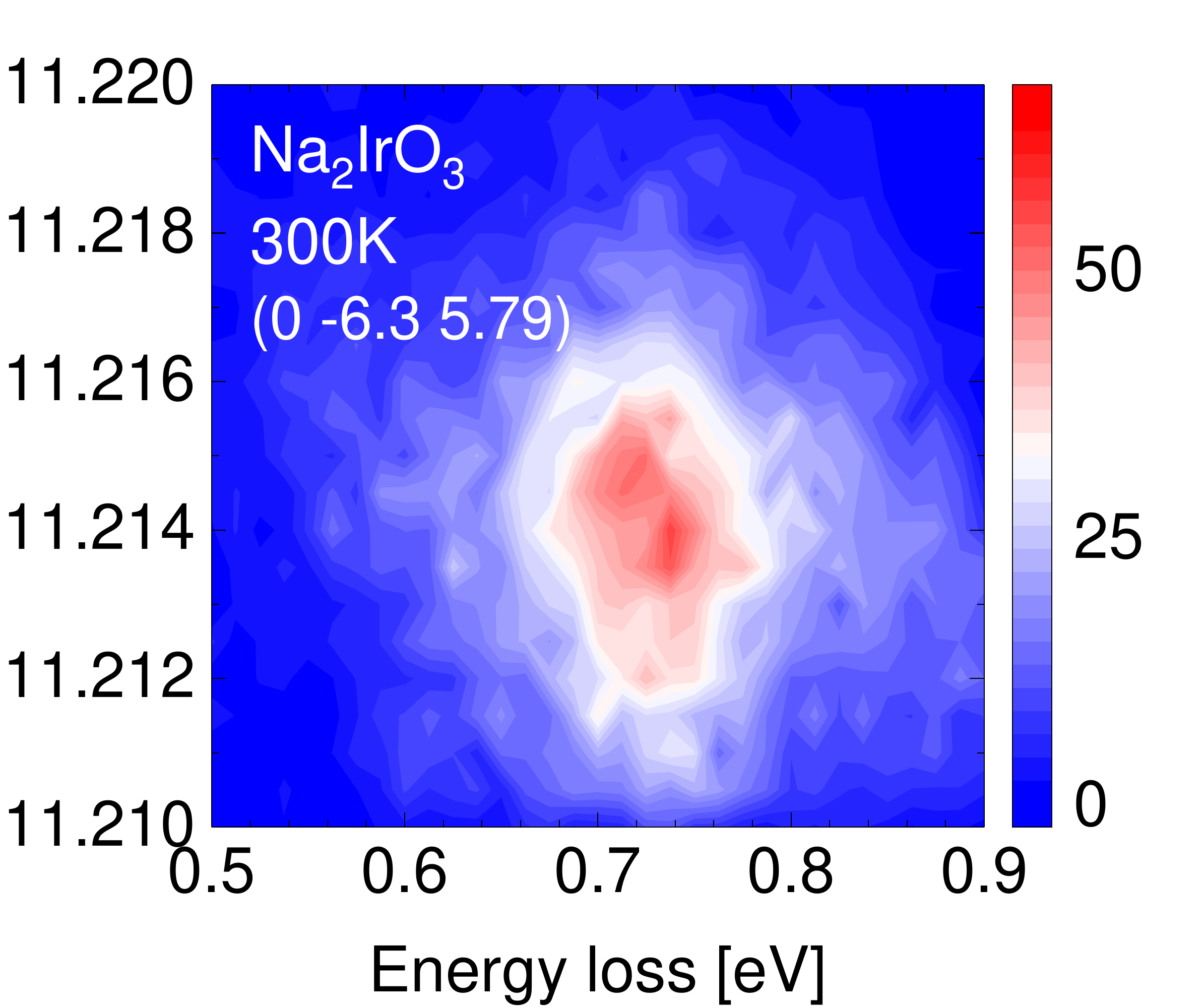}
	\caption{\textbf{Resonance maps.} With RIXS being a resonant technique, the intensity strongly depends on 
		the incident energy which here is varied across the Ir $L_3$ edge. Data were measured for \liir{} 
		at 5\,K (top) and \nair{} at 5\,K (middle) and 300\,K (bottom). 
		Left: low-energy magnetic excitations. Right: intra-$t_{2g}$ excitations to the $j$\,=\,$3/2$ state. 
		All panels show a pronounced $t_{2g}$ resonance at $E_{\rm in}$\,=\,11.2145\,keV but no resonance 
		at the $e_g$ resonance energy $E_{\rm in}$\,=\,11.218\,keV.\@ 
		The elastic line is suppressed by choosing $2\theta$\,$\approx$\,$90^\circ$. 
		Cuts through the resonance maps are plotted in Figs.\ \ref{fig_resonance} and \ref{fig_resonance_cuts}. 
	} 
	\label{fig_resonance_maps}
\end{figure}

In RIXS at the Ir $L_3$ edge, magnetic excitations can be distinguished from phonons by their resonance behavior. 
Magnetic spin-flip excitations and orbital excitations are created in a \textit{direct} RIXS process. 
Starting from a local $j$\,=\,1/2 ground state with a $2p^6\,5d^5$ configuration in which the five $5d$ 
electrons occupy the $t_{2g}$ shell, the incident photon energy of 11.2145\,keV is tuned to resonantly enhance 
the absorption process that promotes a $2p$ core electron to the $5d$ $t_{2g}$ level, 
resulting in a $2p^5\,t_{2g}^6$ intermediate state. 
Within a few femtoseconds, this intermediate state relaxes to the final state by photon emission, 
resulting in an excited $2p^6\,t_{2g}^{5*}$ state that corresponds to a magnetic or orbital excitation. 
The resonance strongly boosts the RIXS intensity of these excitations. 
In contrast, a phonon excitation is an \textit{indirect} RIXS process which arises due to the dynamics 
in the short-lived intermediate state, i.e., the lattice distorts to screen the core hole before 
the electronic system relaxes from the intermediate state to the $j$\,=\,1/2 ground state. 
Typically, the cross section for direct RIXS is much larger \cite{Ament11RMP}. 
Phonons are expected to be particularly weak in RIXS if the core hole in the intermediate state is 
well screened \cite{Ament11RMP}, as commonly assumed for the Ir $2p^5 t_{2g}^6$ intermediate state 
with localized $t_{2g}$ states. This explains why no detectable phonon contribution was observed thus far 
in $L$-edge RIXS data of other Mott-insulating $5d^5$ iridates 
\cite{Kim12,Liu12,KimNatComm14,Gretarsson13PRL,Gretarsson13PRB,MorettiCaIrO3,Gretarsson16,Pincini17,MorettiSr3,Lu17,Rossi17,Revelli19,Revelli19a}. 
A significant phonon contribution requires a more delocalized intermediate state, in which case the core hole 
is screened by adjacent ions and the lattice distorts. 
Since the $e_g$ orbitals are both, more delocalized and more strongly coupled to the lattice, 
the intensity of a hypothetical phonon peak is expected to be enhanced at the $e_g$ resonance \cite{Moser15}, 
which occurs at about $E_{\rm in}$\,=\,11.218\,keV in the $5d^5$ iridates \cite{Lefrancois16}. 
The 3.5\,eV difference in resonance energy between $t_{2g}$ and $e_g$ levels corresponds to the 
cubic crystal field splitting.

\begin{figure}[t]
	\centering
	\includegraphics[width=0.95\columnwidth]{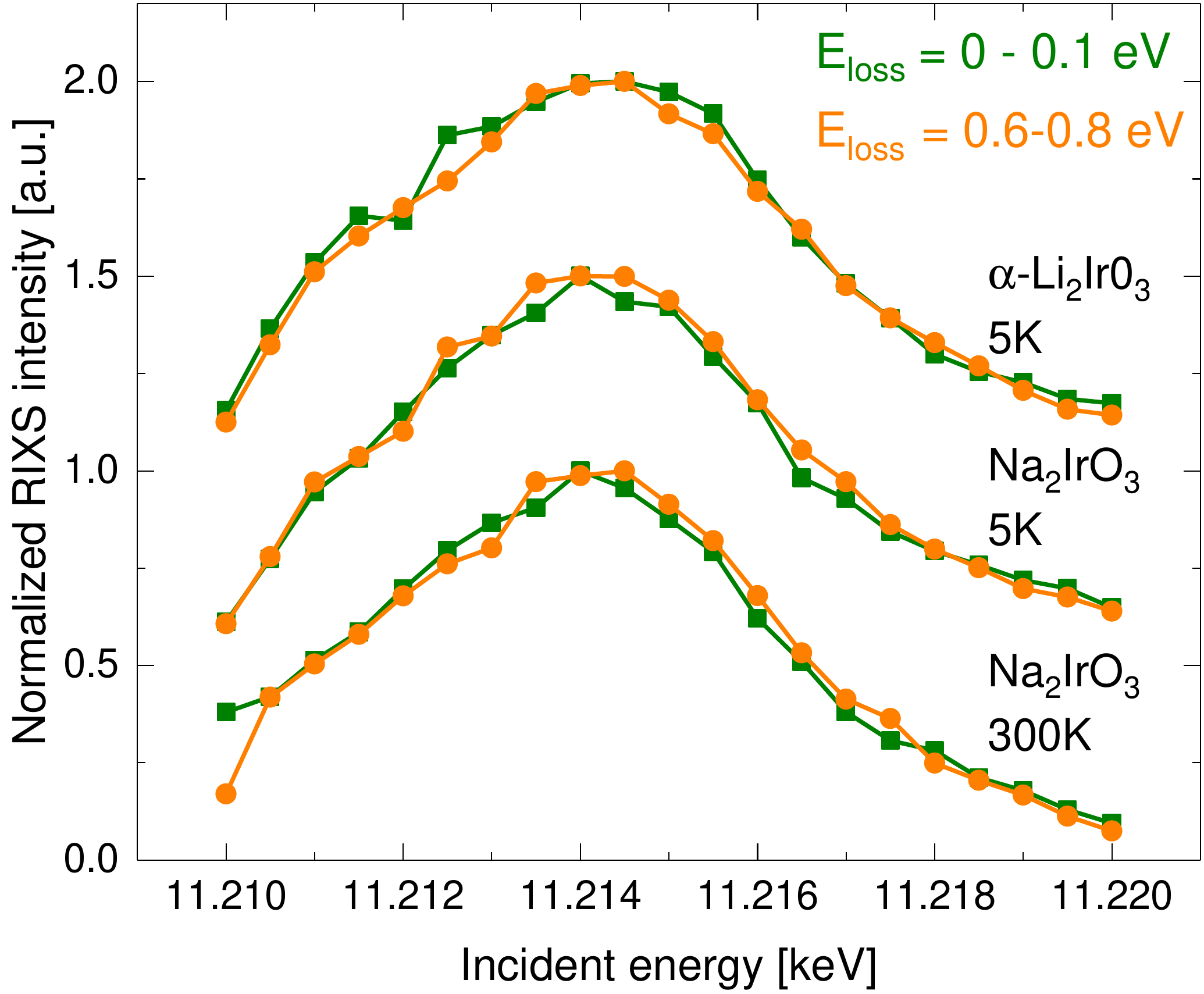}
	\caption{\textbf{Constant-loss cuts through the resonance maps shown in Fig.\ \ref{fig_resonance_maps}.} 
	Data were normalized to the maximum value and are offset for clarity. 	
	In both compounds, the low-energy excitations (green) show the same resonance behavior as the intra-$t_{2g}$ 
	excitations around 0.7\,eV (orange), exhibiting a pronounced resonance peak at 11.2145\,keV.\@ 
	This demonstrates the direct RIXS character of the excitation process and establishes the 
	magnetic character of the low-energy continuum, both at 5\,K and at 300\,K. 
} 
	\label{fig_resonance_cuts}
\end{figure}

The resonance behavior thus provides a litmus test for the magnetic character of excitations. In the iridates, 
magnetic excitations and intra-$t_{2g}$ orbital excitations both resonate at $E_{\rm in}$\,=\,11.2145\,keV, 
whereas the RIXS intensity of phonons is expected to show a different resonance behavior. 
In Mott-insulating \liir{} and \nair, intra-$t_{2g}$ excitations set in at 0.4\,eV with a peak at 
about 0.7\,eV for spin-orbital excitations to the $j$\,=\,3/2 state \cite{Gretarsson13PRL}, 
see Fig.\ \ref{fig_resonance}. 
The resonance maps depicted in Fig.\ \ref{fig_resonance_maps} show the same resonance behavior for the 
RIXS continuum below 0.1\,eV and the intra-$t_{2g}$ excitations above 0.4\,eV, both resonating at 
$E_{\rm in}$\,=\,11.2145\,keV.\@ 
We emphasize that an $e_g$ resonance at 11.218\,keV is not observed.

The experimental result that all excitations in the measured energy range show the {\em same} resonance behavior 
is particularly obvious from cuts through the resonance maps. Normalized constant-loss cuts below 0.1\,eV 
and around 0.6 to 0.8\,eV agree very well with each other, showing a maximum resonance enhancement at 
11.2145\,keV independent of the energy loss, see Fig.\ \ref{fig_resonance_cuts}.  
Cuts for constant $E_{\rm in}$\,=\,11.2145\,keV and 11.218\,keV are compared in Fig.\ \ref{fig_resonance}. 
Also these spectra fall on top of each other after normalization. 
For the normalization, the data for $E_{\rm in}$\,=\,11.218\,keV were scaled up by a factor of  2.8, which 
compensates the detuning of $E_{\rm in}$ away from the $t_{2g}$ resonance. 
This scaling behavior proves a common electronic intra-$t_{2g}$ character of the \textit{entire} excitation 
spectrum below 1\,eV, where `electronic' refers to magnetic and orbital degrees of freedom at 
low and high energies, respectively. Most noteworthy, this still holds true at 300\,K, 
see the bottom panel of Fig.\ \ref{fig_resonance}. 
Thus, we conclude that a phonon interpretation of the low-energy continuum can be firmly ruled out.

\begin{figure}[t]
	\centering
	\includegraphics[width=1.0\columnwidth]{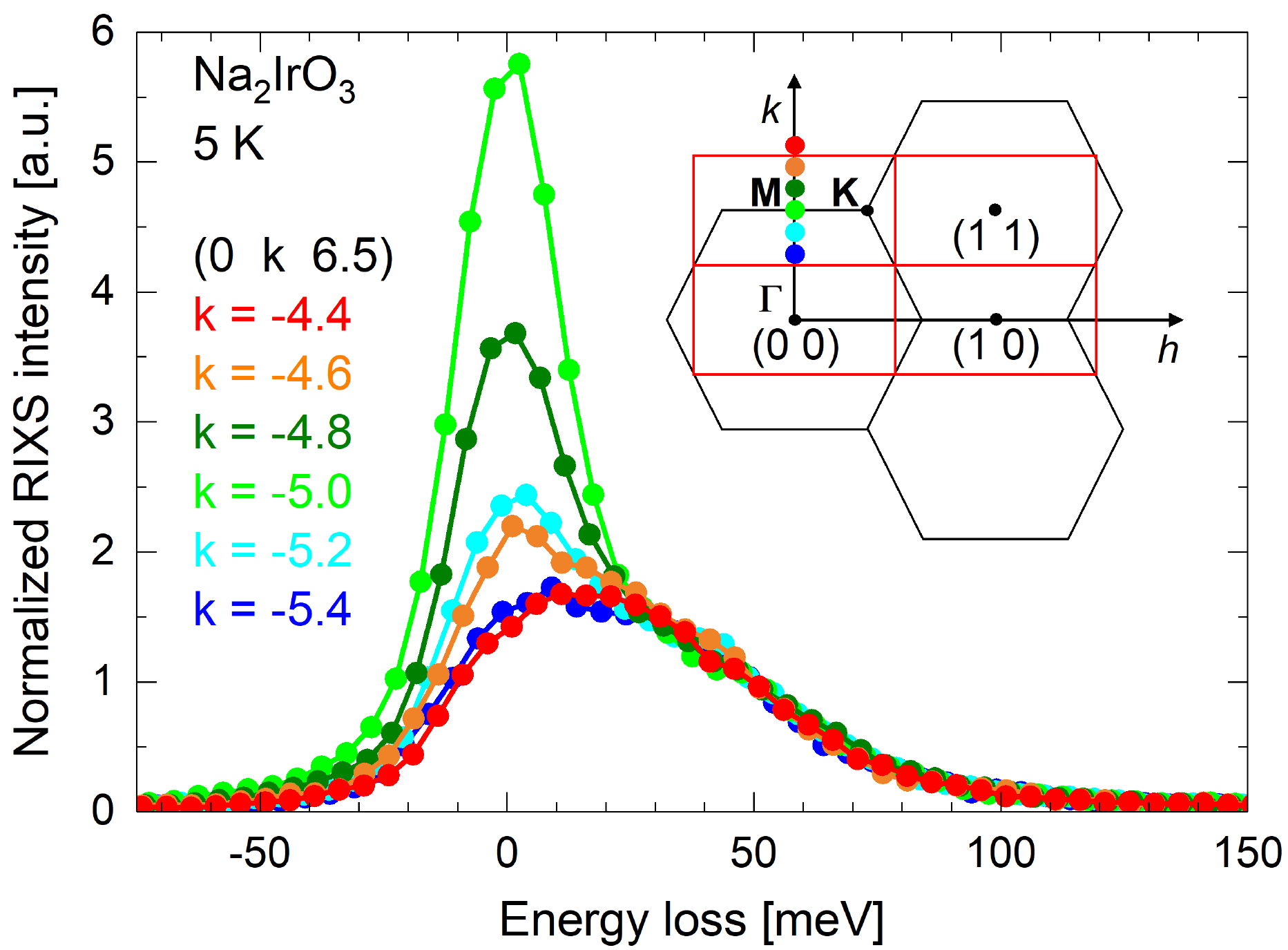}
	\caption{\textbf{Momentum dependence of normalized RIXS spectra of \nair.} 
	Data were normalized by the intensity at 50\,meV energy loss. 
	Above about 20\,meV, the normalized spectra fall on top of each other, showing that concerning the continuum, 
	only the intensity changes as a function of $\mathbf{q}$. 
	Inset: Black lines denote the 2D Brillouin zone of the honeycomb lattice with the high-symmetry points 
	$\Gamma$, $K$, and $M$. Red rectangles depict 2D cuts of the Brillouin zone of the monoclinic lattice.
	The colors of the symbols denote the $\mathbf{q}$ values equivalent to the ones studied in the main figure.
} 
	\label{fig_Na_spectra_scaled}
\end{figure}

\subsection{Nearest-neighbor spin-spin correlations}

Having established the magnetic character of the continuum, we now address the $\mathbf{q}$-dependent behavior. 
RIXS spectra of \nair{} covering the range from (0 -4.4 6.5) to \mbox{(0 -5.4 6.5)} are depicted in Fig.\ \ref{fig_Na_spectra_scaled}. Note that (0 -5 6.5) corresponds to a magnetic Bragg peak \cite{Ye12}. 
The data in Fig.\ \ref{fig_Na_spectra_scaled} were normalized to the RIXS intensity at an energy loss of 
50\,meV.\@ The excellent agreement obtained above some 20\,meV shows that the RIXS intensity 
of the magnetic continuum depends on $\mathbf{q}$ while the peak energy and width are hardly affected. 
The $\mathbf{q}$-dependent intensity of the continuum corresponds to the dynamical structure factor 
$S(\mathbf{q},\omega)$, which allows us to measure dynamical spin-spin correlations in real space
(see also the \textit{Discussion} section).

\begin{figure}[t]
	\centering
	\includegraphics[width=1\columnwidth]{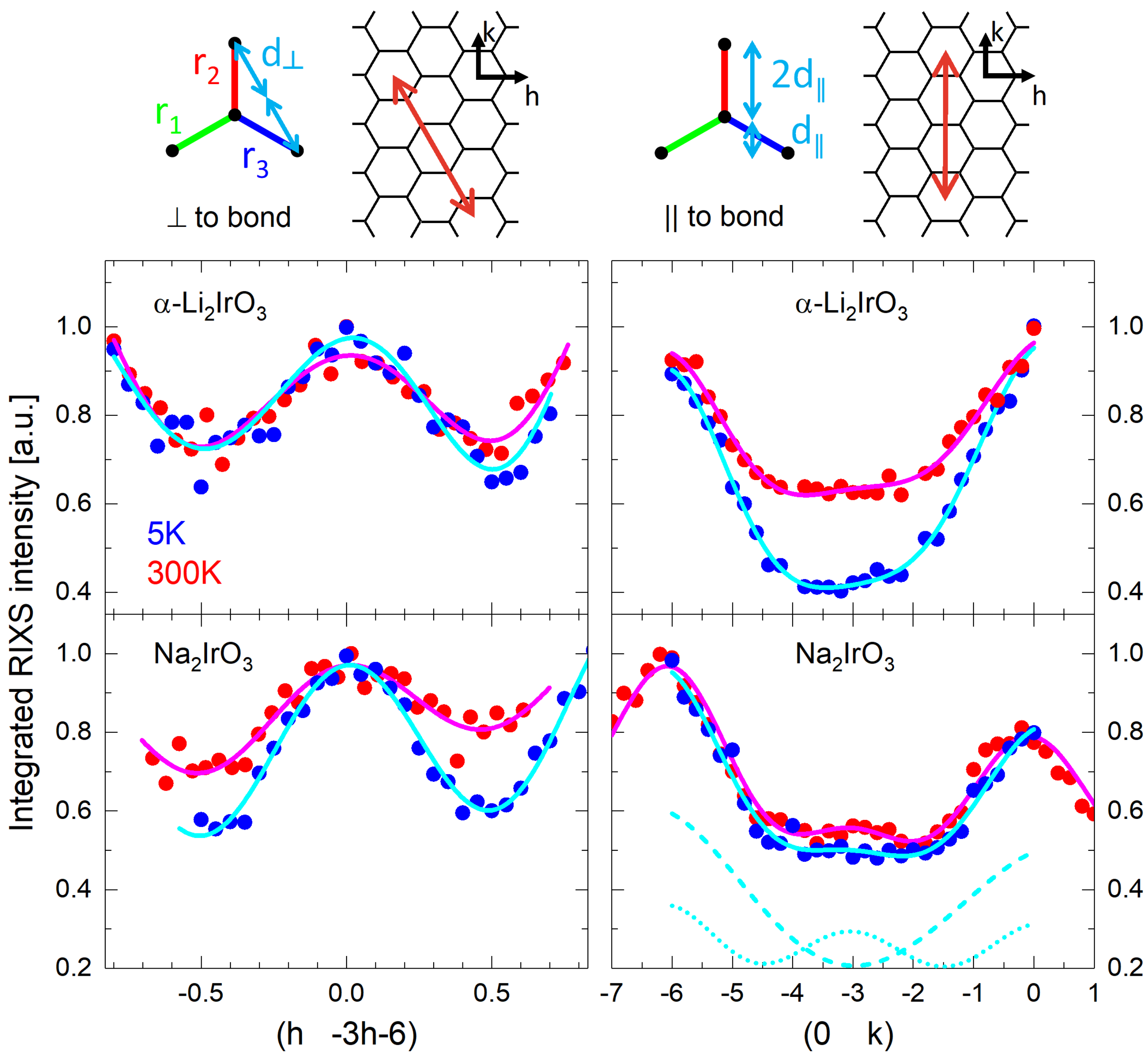}
	\caption{\textbf{RIXS interference patterns showing only nearest-neighbor spin-spin correlations.} 
	All panels depict the normalized RIXS intensity as a function of $\mathbf{q}$, integrated from 
	30 to 150\,meV.\@ 
	Data were corrected for self-absorption effects \cite{Minola15}.
	Solid lines: sinusoidal fits assuming that dynamical spin-spin correlations are restricted 
	to nearest neighbors. 
	Top: \liir. Bottom: \nair. 	
	Left (right): data for $\mathbf{q}$ along $\Gamma$-$K$ ($\Gamma$-$M$), 
	as sketched on top, where the red arrows mark the range of $\mathbf{q}$ covered in the panels below. 
	Along $\Gamma$-$K$ (left), the projection of $\mathbf{q}$ on the Ir-Ir bonds yields a single 
	sinusoidal modulation. Along $\Gamma$-$M$, nearest-neighbor correlations along the three different 
	bonds yield a superposition of two modulation periods, see Eq.\ \eqref{eq:I},
	as observed experimentally. Dashed and dotted lines in the bottom right panel exemplary show the 
	two individual contributions to the fit at 5\,K.\@ 
	The interference patterns are remarkably robust against temperature. 
	Sketches on top show the three Ir-Ir bonds $\mathbf{r}_i$ as well as several Brillouin zones 
	of the reciprocal lattice (black).
} 
	\label{fig_all_k_scans}
\end{figure}

We address $S(\mathbf{q},\omega)$ by measuring the $\mathbf{q}$-dependent intensity 
modulation of the magnetic continuum above 30\,meV for different directions of $\mathbf{q}$, 
see Fig.\ \ref{fig_all_k_scans}. The data cover several Brillouin zones, both at 5\,K and 300\,K.\@ 
Note that the honeycomb lattice has three different Ir-Ir bonds described by the real-space vectors 
$\mathbf{r}_{1/3}$\,=\,($\pm a/2$, $b/6$, 0) and $\mathbf{r}_2$\,=\,(0, $b/3$, 0), 
as sketched by the green, red, and blue bonds on top of Fig.\ \ref{fig_all_k_scans}.
The left panels of Fig.\ \ref{fig_all_k_scans} show data for $\mathbf{q}\perp \mathbf{r}_1$ with 
$\mathbf{q}\cdot\mathbf{r}_2$\,=\,$\mathbf{q}\cdot\mathbf{r}_3$ ($\Gamma$-$K$ direction, 
cf.\ inset of Fig.\ \ref{fig_Na_spectra_scaled}). In this direction, we find a pronounced 
sinusoidal intensity modulation $\cos^2(\mathbf{q}\cdot\mathbf{r}_2/2)$ in both compounds.
The modulation period corresponds to nearest-neighbor correlations. 
This behavior strongly deviates from the expectations for magnons in a long-range ordered state, 
as reported in $L$-edge RIXS on the square-lattice Heisenberg magnet Sr$_2$IrO$_4$ \cite{Kim12}. 
Recently, such a sinusoidal modulation of the Ir $L$-edge RIXS intensity was observed for Ir dimers in 
Ba$_3$CeIr$_2$O$_9$ where it was discussed as an inelastic incarnation of Young's double slit 
experiment \cite{Revelli19}. 
Applied to the magnetic continuum, it shows that a given excited state can only be reached by 
spin-flip RIXS processes on either of the two Ir sites of a given bond. In other words, 
dynamical spin-spin correlations for the magnetic continuum are restricted to nearest neighbors. 
This is a key feature of the Kitaev model, in which the three different spin-flip channels are independent 
and do not interfere with each other \cite{Halasz16}. Flipping the $x$ component of the spin creates 
an excitation on an $x$ bond, while a flip of the $y$ component yields an orthogonal final state living 
on a $y$ bond. 
Similar results were reported for \rucl{} based on inelastic neutron scattering \cite{Banerjee17}. 
Remarkably, \liir{} and \nair{} show very similar behavior, despite their different magnetic ordering 
patterns \cite{Winter17rev}.

\begin{figure}[t]
	\centering
	\includegraphics[width=1\columnwidth]{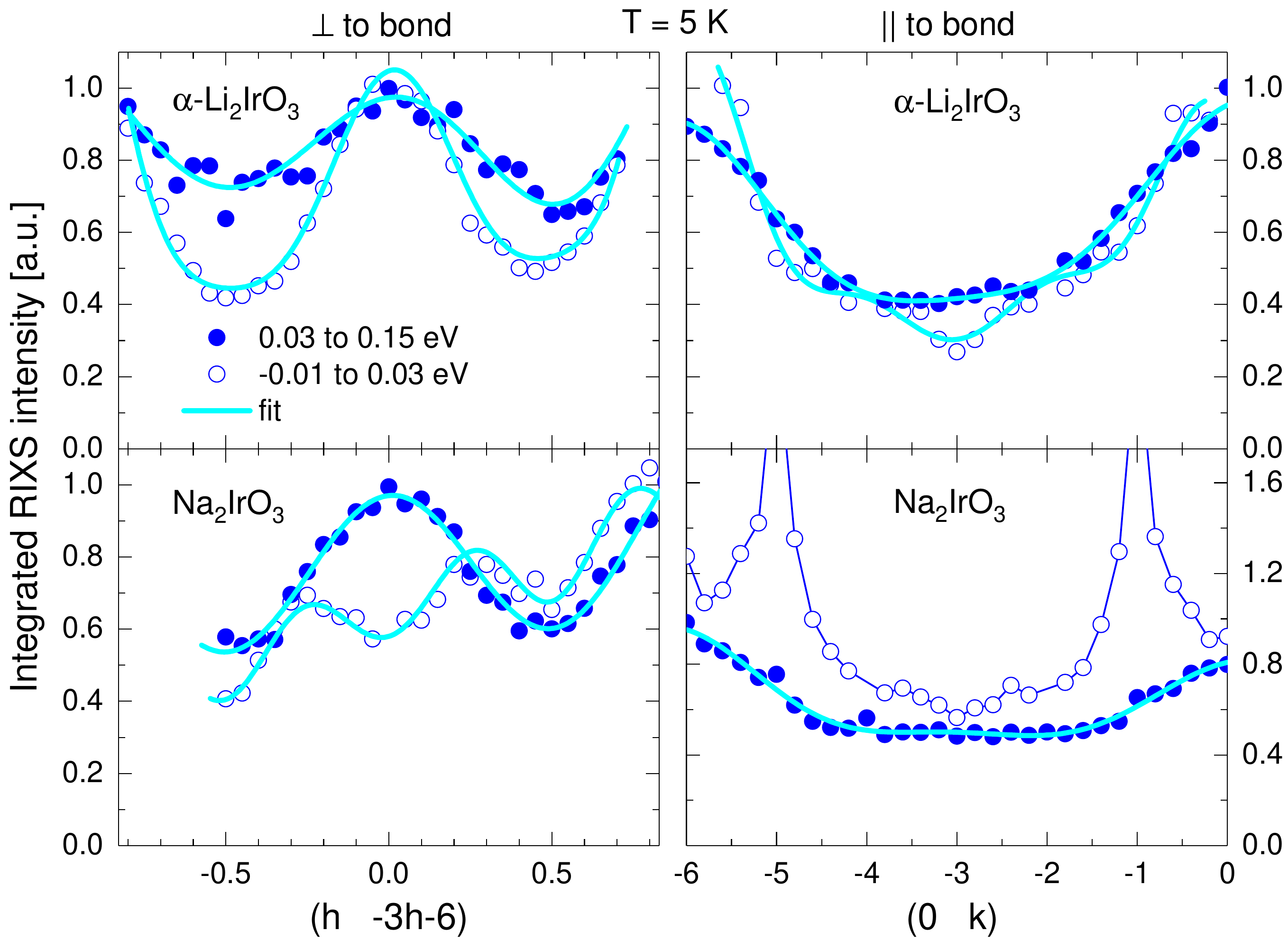}
	\caption{\textbf{RIXS interference patterns in the long-range magnetically ordered state at 5\,K.} 
	Full symbols: same data as in Fig.\ \ref{fig_all_k_scans}, showing the normalized RIXS intensity 
	integrated from 30 to 150\,meV.\@ Open symbols show corresponding low-energy data integrated from 
	-10 to +30\,meV, revealing differences between the two compounds that can be attributed to the 
	different magnetic ordering patterns. In contrast, the behavior at high energy (full symbols) is identical. 
	Top: \liir. Bottom: \nair. 
	Left: $\Gamma$-$K$ direction (perpendicular to an Ir-Ir bond, see Fig.\ \ref{fig_all_k_scans}). 
	Right: $\Gamma$-$M$ direction (parallel to an Ir-Ir bond). 
	Data were corrected for self-absorption effects \cite{Minola15}.
} 
	\label{fig_all_k_scans_5K}
\end{figure}

Besides the sinusoidal modulation, the RIXS intensity shows a slow variation with $\mathbf{q}$. 
Note that a change of $\mathbf{q}$ is unavoidably accompanied by a change of the experimental scattering 
geometry, which affects the x-ray polarization and the matrix elements. 
This gives rise to deviations from a pure sinusoidal behavior.
We emphasize that spin-spin correlations beyond nearest neighbors would give rise to 
a \textit{shorter} modulation period, for which we do not find any evidence. Based on the noise level of 
the data, we estimate the upper limit for such a further-neighbor contribution with a shorter 
modulation period to 10\%.

The restriction of dynamical spin-spin correlations to nearest neighbors is corroborated by the behavior 
observed for the $\Gamma$-$M$ direction, see right panels of Fig.\ \ref{fig_all_k_scans}. 
With $\mathbf{q}\parallel \mathbf{r}_2$ and 
$\mathbf{q}\cdot \mathbf{r}_1$\,=\,$\mathbf{q}\cdot \mathbf{r}_3$\,=\,$(\mathbf{q}\cdot \mathbf{r}_2)/2$, 
the nearest-neighbor correlations on the three different bonds give rise to an incoherent sum of 
two sinusoidal modulations, 
\begin{equation}
I(\mathbf{q}) \propto I_2 \cos^2(\mathbf{q}\cdot\mathbf{r}_2/2) + I_{1/3} \cos^2(\mathbf{q}\cdot\mathbf{r}_{1/3}/2) 
\, , 
\label{eq:I}
\end{equation}
where the period of the latter is twice the period of the former due to the different projections 
$\mathbf{q}\cdot\mathbf{r}_i$, see dashed and dotted lines in the lower right panel of Fig.\ \ref{fig_all_k_scans}. 
Most importantly, we do not find, within our experimental resolution,
any evidence for further modulation periods beyond nearest neighbors. 
The experiment roughly shows $I_{1/3} \approx 2 I_2$, which reflects the fact that the two bonds $\mathbf{r}_1$ 
and $\mathbf{r}_3$ contribute to the modulation amplitude $I_{1/3}$. 
Additionally, the prefactors depend on the x-ray polarization and the scattering geometry, 
both changing with $\mathbf{q}$ in our measurements. As discussed above for the $\Gamma$-$K$ direction, 
this causes deviations from a pure sinusoidal behavior.

The $\mathbf{q}$-dependent intensity modulation is very robust against temperature. 
The main change at 300\,K is a reduction of the modulation amplitude, see Fig.\ \ref{fig_all_k_scans}. 
This can be rationalized by the energy scale of the observed excitations, which is comparable to 
$k_B \,300\,K \approx 26\,$meV.\@ 
It is thus not surprising that nearest-neighbor correlations are still strong at 300\,K.\@

Our central experimental finding is that the magnetic continuum shows only nearest-neighbor 
spin-spin correlations even in the long-range magnetically ordered state at 5\,K.\@
This behavior strongly deviates from the expectations for magnons, 
which are supposed to reflect longer-range spin-spin correlations at lower energy.
Indeed, for integration below 30\,meV at 5\,K, both \nair{} and \liir{} show additional modulation periods 
which correspond to spin-spin correlations beyond nearest neighbors, 
see open symbols in Fig.\ \ref{fig_all_k_scans_5K}. 
Due to the different magnetic ordering patterns, the behavior of the two compounds differs at low energy 
and for temperatures below $T_N$. In \nair, the intensity for instance strongly increases 
in the vicinity of (0 -1 6.5), a magnetic Bragg peak reflecting zigzag order. 
The incommensurate magnetic order of \liir{} gives rise to a different dynamical structure 
factor $S(\mathbf{q},\omega)$. 
These differences observed at low energy corroborate the notion that the common high-energy 
continuum is not related to long-range magnetic order, as sketched in Fig.\ \ref{fig_sketch_phase}.

\section{Discussion}

Our RIXS study of \nair{} and \liir{} in the high-temperature regime above their respective magnetic 
ordering transitions as well as the high-energy regime of their magnetically ordered low-temperature states 
has revealed many atypical features that allow to clearly rule out that this high-energy and high-temperature 
RIXS response can be interpreted as arising from a conventional, featureless paramagnetic state. 
In fact, taken together these experimentally observed features -- a broad continuum of magnetic excitations, its persistence to high temperatures, and dominant nearest-neighbor spin-spin correlations -- can be consistently explained, as we will argue in the following,
by assuming that \nair{} and \liir{} are both in close proximity to a Kitaev spin liquid. To do so, we will review the RIXS 
response expected for the pure Kitaev model and discuss how this response is modified in the presence of additional
magnetic interactions \cite{Winter16}, i.e.~generic diagonal and off-diagonal couplings expected to be present in these spin-orbit 
entangled materials \cite{Rau2016}.

 \begin{figure}[b]
	\centering
	\includegraphics[width=1.0\columnwidth]{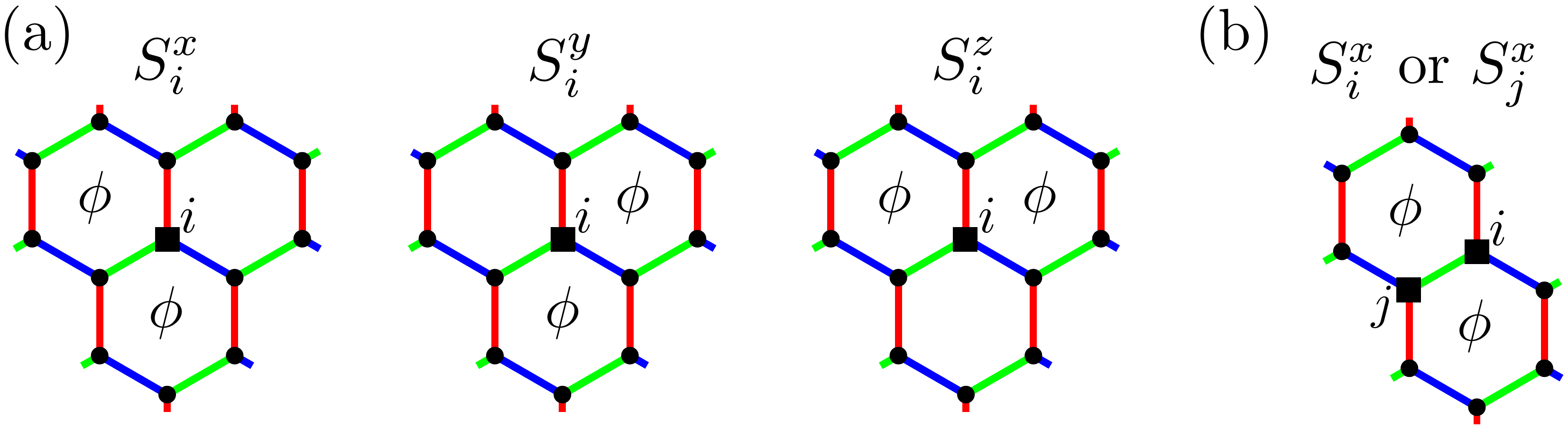}
	\caption{{\bf Flux excitations in the Kitaev model.}
	 In the Kitaev model, spin-flip RIXS on site $i$ creates two flux excitations, so-called visons, 
	on adjacent plaquettes. 
	(a) For excitation on site $i$, there are three orthogonal excited states, each corresponds to 
	a different spin-flip channel $S_i^\alpha$ \cite{Halasz16}. 
	(b) Each excited state can  be reached only from the two sites $i$ and $j$ on the bond connecting 
	the respective plaquettes. 
	}
\label{fig_fluxes}
\end{figure}

\noindent \textit{RIXS response of pure Kitaev model.--}
For the {\em pure} Kitaev model, the RIXS response was first elucidated by Hal\'asz, Perkins, and van den Brink \cite{Halasz16}, where they showed that, with a $t_{2g}^6$ intermediate state, the response naturally decomposes into four \textit{independent} RIXS channels. 
The single spin-conserving RIXS channel corresponds to an indirect RIXS process 
and is thus much weaker than the remaining three {\em spin-flip} RIXS channels, which correspond to direct processes. As a result, it is only these spin-flip RIXS channels that are of relevance to the discussion here. The three different channels correspond to the three diagonal components of the dynamical spin structure factor. In the language of the Kitaev model, they excite both mobile Majorana fermions and static flux excitations, so-called visons. For example, the operator $S_i^x$ creates propagating Majoranas, as well as two visons in adjacent plaquettes, as illustrated in Fig.~\ref{fig_fluxes}(a). 
Since the visons are gapped, this translates to a gapped RIXS response, with a {\em sharp peak at the $\Gamma$ point} at the two-flux gap of the order of ten percent of the Kitaev coupling $K_1$, 
followed by a {\em broad continuum}, reflecting the existence of fractionalized excitations (itinerant Majorana fermions and static visons).

\textit{Broad continuum of excitations.--}
Experimentally, the total spin-flip RIXS spectrum shows only a broad magnetic excitation continuum at the $\Gamma$ point, peaking at about 15\,meV in \liir{} and 35\,meV in \nair. The expectation of the pure Kitaev model, a dominant sharp peak at $\hbar\omega\,\sim\,0.1\,|K_1|$ followed by a broad continuum, must be modified to include (i) the anisotropy of the Kitaev coupling on different bonds \cite{Winter16}, 
(ii) the inclusion of additional exchange interactions such as Heisenberg and off-diagonal, bond-directional terms (such as the so-called $\Gamma$-exchange) \cite{Rau2016}, 
and (iii) finite temperature. Taken together, all three effects lead to a significant broadening of the dominant peak and a shift to higher energies \cite{Knolle14,Knolle18,Yoshitake16,Yoshitake17a,Samarakoon17}.
For both \nair{} and \liir{} the anisotropy of $K_1$ and the existence of additional exchange interactions 
are well established \cite{Winter16}. Unfortunately, inclusion of such terms breaks the integrability of the model, providing a challenge to quantitative theoretical modeling. However, an augmented parton mean-field theory from Knolle, Bhattacharjee, and Moessner \cite{Knolle18} found that 
the dominant low-energy peak does indeed broaden substantially and shift to significantly higher energy upon inclusion of the additional relevant exchange interactions. Without detailed knowledge of the minimal model for the materials at hand, one can only estimate that the peak energy should be comparable to the energy scale of the Kitaev coupling. The ratio of the energy at which the peak occurs for \nair{} to that for \liir{} is roughly a factor of two. Form this, we are led to infer a similar ratio of their respective Kitaev couplings. Indeed, this is in good agreement with {\em ab-initio} estimates \cite{Winter16,Katukuri14,Nishimoto16,Yamaji14}, which yield ferromagnetic $K_1$ in the range -15 to -35\,meV for \nair{} and -4  to -13\,meV 
for \liir{}. The broad excitation continuum observed is thus consistent with the expectation for an extended Kitaev model.

\begin{figure}[t]
	\centering
	\includegraphics[width=0.9\columnwidth]{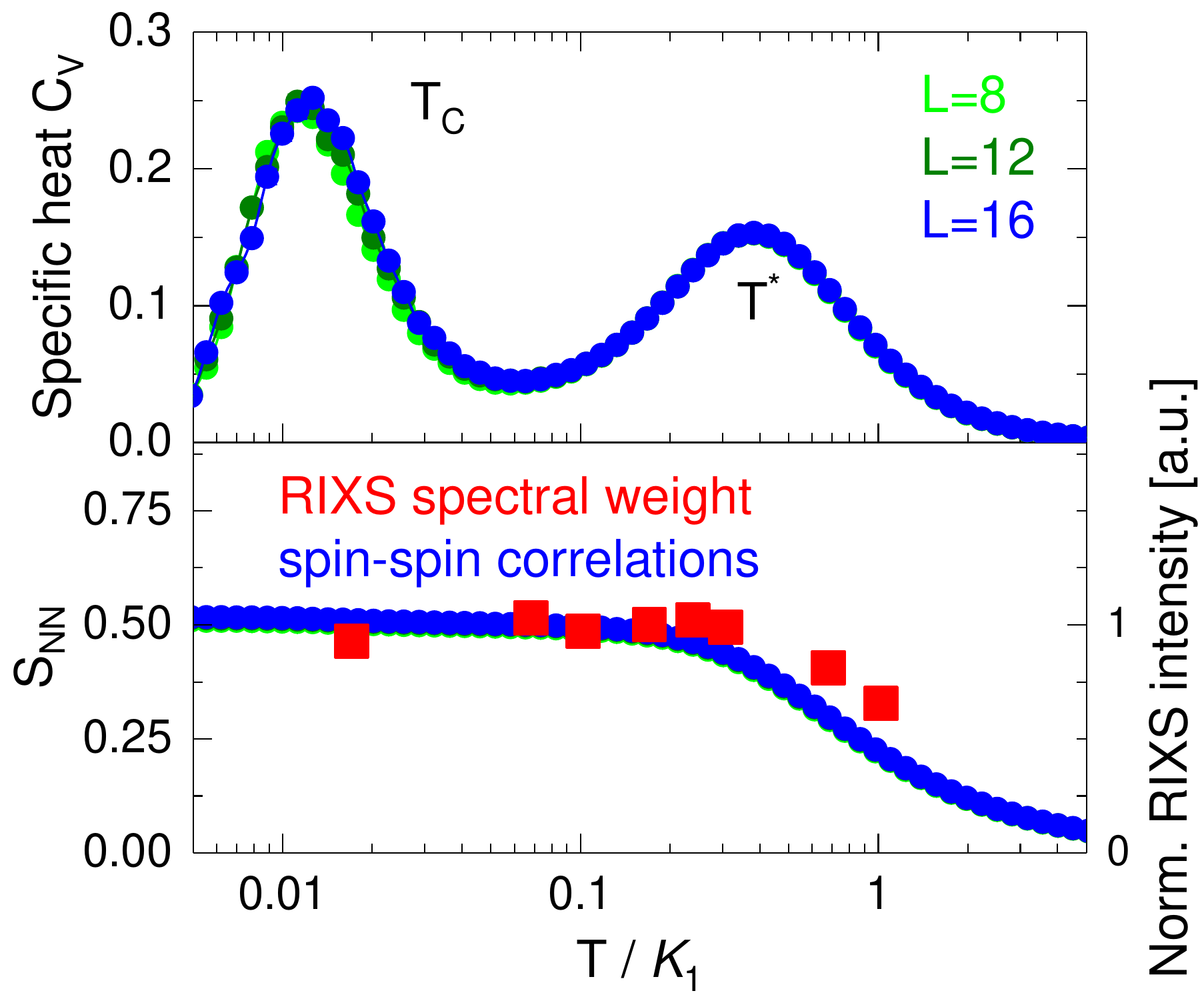}
	\caption{{\bf Thermodynamic signatures of Kitaev physics and temperature dependence of spin-spin correlations.} 
		Top: The specific heat of the pure Kitaev model exhibits a characteristic two-peak structure, 
		with a crossover at $T^* \! \sim \! K_1$ to a disordered Kitaev spin liquid regime and a low-temperature
		transition at $T_c \approx K_1/100$ to the Kitaev spin liquid phase (see also Fig.\ \ref{fig_sketch_phase}). 
		Bottom: The nearest neighbor spin-spin correlations remain nearly constant up to the high-temperature
		crossover scale. 
		Numerical results for the pure Kitaev model from quantum Monte Carlo simulations (blue) 
		are shown in comparison to the temperature dependence of the RIXS intensity of \nair{} integrated above 30\,meV 
		(red,	cf.\ Fig.~\ref{fig_Na_T}). 
		To set the temperature scale for the comparison, we used $T^*$\,=\,110\,K as determined from 
		specific heat data \cite{Mehlawat17}. } 
	\label{fig_Szz}
\end{figure}

\textit{Temperature dependence of continuum.--}
With increasing temperature, the shape of the experimentally observed broad continuum remains largely unchanged, with only a decrease in overall intensity. 
This decrease is more pronounced for \liir{}, consistent with the overall lower energy scale of the peak. Such a temperature dependence is qualitatively consistent with numerical calculations, where, at temperatures above the flux gap, the broad continuum displays only a smooth weakening in intensity as temperature increases \cite{Yamaji18,Yoshitake16,Samarakoon17,Suzuki18}.
Since the magnetic continuum reveals only nearest-neighbor correlations, see Fig.~\ref{fig_all_k_scans}, we can quantitatively compare the  integrated RIXS intensity of the continuum with the (static) nearest-neighbor correlations of the pure Kitaev model, calculated through numerically exact quantum Monte Carlo simulations \cite{Nasu2014}.  
Note that we consider static correlations in our numerics, i.e.~the dynamical structure factor integrated 
over {\em all} energies, while in experiment the data was integrated above a cutoff of 30\,meV.\@  
Such a comparison is shown in Fig.~\ref{fig_Szz}, where we have rescaled the temperature scale,
using the experimentally determined  specific heat data of \nair{} \cite{Mehlawat17} which locates the 
high-temperature crossover in the specific heat (see upper panel in Fig.~\ref{fig_Szz}) at $T^*$\,=\,110\,K.\@ 
After this rescaling, the two data sets bear a striking resemblance -- both traces of the nearest-neighbor correlations remain nearly constant up to the order of the high-temperature crossover scale and drop only then, indicating a transition to a featureless, paramagnetic state at temperatures which are one order of magnitude higher than the experimentally observed magnetic ordering scale.

We note that an alternate scenario for the origin of the broad continuum in \rucl{} has recently been 
put forward by Winter \textit{et al}.\  \cite{Winter17}, which attributes it to {\em magnon decay} within a multi-magnon continuum. They showed that large anisotropic couplings in spin-orbit coupled magnets can generically lead to substantial two-magnon decay, giving rise to a broad, incoherent 
excitation continuum in the long-range ordered state. 
This raises the question of whether the magnetically ordered state or the proximate spin liquid are the better starting point for 
the description of the excitations in \nair{} and \liir{}. In our experimental RIXS data, one of the most striking observations is that the excitation continuum persists up to temperatures of roughly 20\,$T_N$, far above the ordered state and magnon modes. 
Consequently, we are led to the conclusion that we cannot explain the existence of a high-temperature 
quasi-2D continuum via the physics of low-temperature 3D magnons and that the scenario of a proximate spin liquid is a better starting point.

\begin{figure}[t]
	\centering
	\includegraphics[width=1.0\columnwidth]{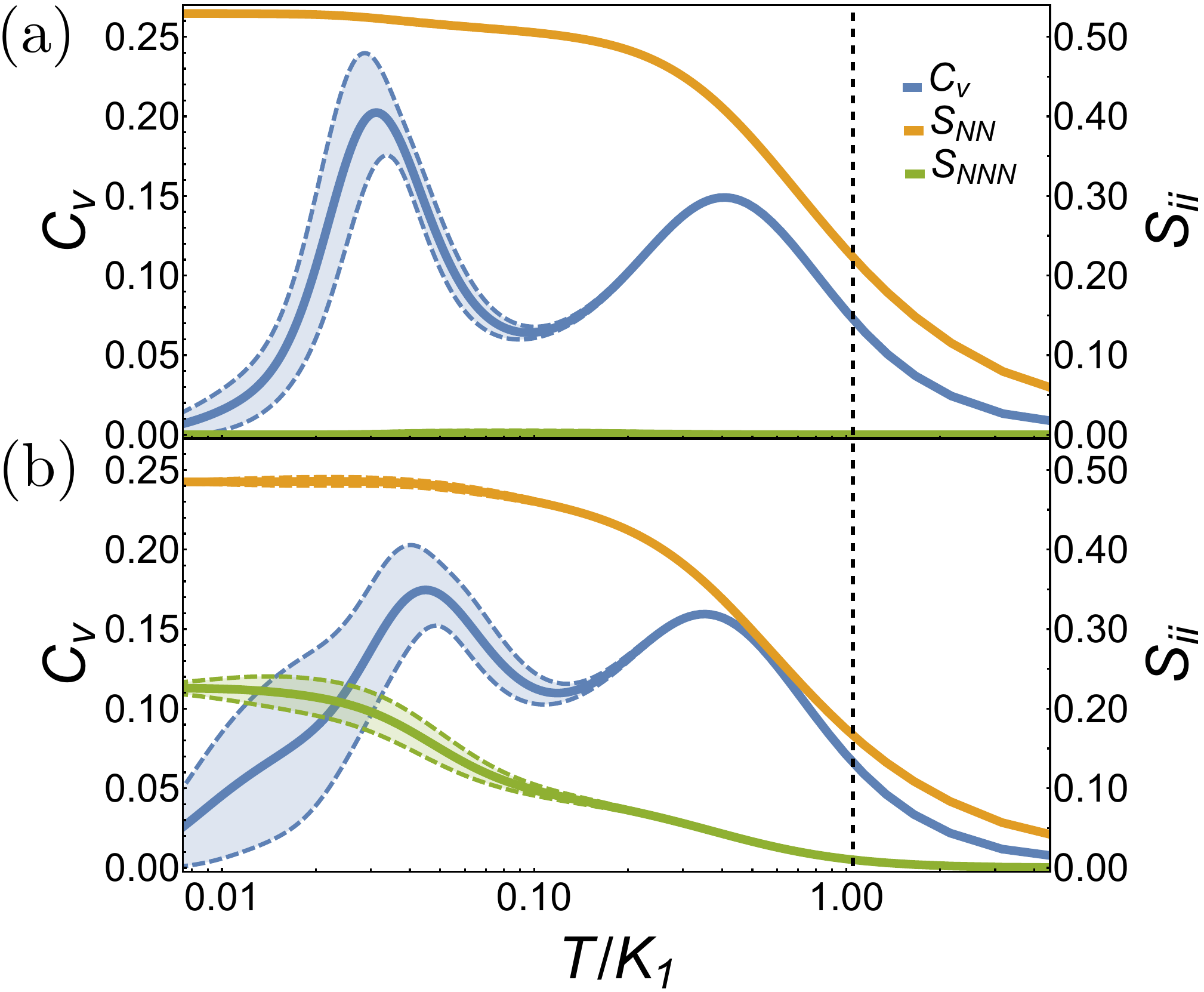}
	\caption{{\bf Temperature dependence of static spin-spin correlations.}
			Comparison of the  nearest (orange) and next-nearest neighbor (green) spin-spin correlations, $S_{NN}$ and $S_{NNN}$, 
			as a function of temperature for (a) the pure Kitaev model and (b) an extended Kitaev model
			augmented by Heisenberg and off-diagonal spin exchanges as proposed \cite{Yamaji14}
			by Yamaji {\em et al}.\ for \nair{}.
			The next-nearest neighbor correlations remain vanishingly small in our exact diagonalization
			studies for the pure Kitaev model. However, in the extended model, they are finite at zero temperature but start to decrease around the
			temperature scale of the low-temperature specific heat peak (blue). The dashed line indicates $300$\,K, using the same scale as in Fig.~\ref{fig_Szz}. 
			The shaded regions indicate the estimated error of the finite-temperature calculation.
}
\label{fig_Szz_temperature}
\end{figure}

 \textit{Nearest-neighbor spin-spin correlations.--}
As discussed above for the pure Kitaev model, spin-flip RIXS creates two local visons in adjacent plaquettes as illustrated in Fig.~\ref{fig_fluxes}. Such a probe of only the nearest-neighbor spin correlations is indeed corroborated by our experimentally observed momentum dependence. 
Note that this observation clearly differs from the response of other spin liquid candidates inspected 
on the Kagome lattice and the triangular lattice which show clear deviations from a pure $\cos^2(qR/2)$ 
behavior \cite{Han12,Shen16} due to finite correlations between next-nearest and further neighbors.
The pronounced sinusoidal shape thus provides strong evidence for the dominant role of Kitaev interactions.

The strict nearest-neighbor spin-spin correlations of the pure Kitaev model are, of course, modified by the inclusion of the necessary additional interactions. These integrability-breaking terms add dynamics to the previously immobile flux excitations, giving rise to correlations beyond nearest neighbors \cite{Yamaji16,Knolle18}. 
To quantitatively illustrate this effect, we plot results from numerical exact diagonalization calculations \cite{TPQS1,TPQS2}, similar to Ref.~\onlinecite{Yamaji16}, for the (static)
nearest and next-nearest neighbor spin-spin correlations in Fig.~\ref{fig_Szz_temperature}.
Shown are data for both, the pure Kitaev model (top) and an extended Kitaev model (bottom)
augmented by Heisenberg and off-diagonal spin exchanges, based on the proposal \cite{Yamaji14} by 
Yamaji {\em et al}.\ for \nair{}:
$J_1$\,=\,$+3.2$, $K_1$\,=\,$-29.4$, $\Gamma'_1$\,=\,$-3.5$, $J_3$\,=\,$+1.7$ in units of meV.\@ 
As can be seen, these static next-nearest neighbor correlations remain negligible at elevated temperatures, 
remaining at a level well below 20\% of the corresponding nearest-neighbor correlations in the temperature region 
corresponding to the upper peak in the specific heat. 
For a detailed discussion of the role of the individual exchange terms on the nearest and next-nearest neighbor spin-spin correlations, see the Appendix.

 \textit{Comparison to \rucl.--}
Phenomenologically, we note that the results presented here are in full analogy
with the inelastic neutron scattering data on the closely
related $j$\,=\,1/2 compound \rucl{} \cite{Banerjee17}. 
Both in \rucl{} and the honeycomb iridates studied here, there is a broad excitation continuum with a maximum of intensity at the $\Gamma$ point at high energies 
which is robust against temperature, persisting up to roughly $20\,T_{\rm N}$. 
Additionally, the neutron data show a nearly sinusoidal modulation along $\Gamma$-$K$ with a period that reflects that spin-spin correlations are predominantly of nearest-neighbor type. 
Similar to our analysis of the RIXS data, this modulation was interpreted as a signature of the 
experimental proximity to the Kitaev spin liquid. 
Note that in our RIXS data of \liir{} and \nair{}, the deviations from a pure nearest-neighbor behavior are even 
smaller than in \rucl.

\subsection*{Conclusions} 
In conclusion, we find that our experimental RIXS probes of the magnetic excitations of the honeycomb iridates \nair{} and \liir{} reveal compelling evidence for the emergence of fractional excitations in a broad temperature and energy regime. Both at temperatures up to an order of magnitude above the magnetic ordering transition as well as for high-energy probes in the magnetically ordered ground state, we find evidence of a broad continuum of magnetic states
which cannot be explained in terms of low-energy magnon modes. 
In addition, a careful analysis of the (spin-flip) RIXS processes reveals that the spin-spin correlations 
are, within our experimental resolution, confined to nearest neighbor sites. While extremely localized in nature, the spin-spin correlations are extremely robust against thermal fluctuations up to temperatures of about 300\,K.\@ 
In combination, these experimental findings prove that both materials must be located in close proximity to a Kitaev spin liquid, which leaves a distinct fingerprint on their high-temperature and high-energy regimes beyond their magnetically ordered ground states.  

We note that the principal findings of our RIXS experiments for the honeycomb iridates \nair{} and \liir{} are in striking analogy to recent inelastic neutron scattering experiments \cite{Banerjee17} on \rucl{}. Their common phenomenology
-- a broad  continuum of magnetic excitations and spin-spin correlations confined to nearest neighbors, which are both stable way beyond their magnetic ordering temperatures -- should be taken as a concrete notion of what constitutes a ``Kitaev material" 
beyond the existence of a dominant bond-directional exchange \cite{Trebst2017}, and makes these compounds
prototypical examples of this new class of materials.

\begin{acknowledgments}
We acknowledge funding from the Deutsche Forschungs\-gemeinschaft (DFG, German Research Foundation) -- Project numbers 277146847, 247310070, and 107745057 -- CRC 1238 (projects A02, B03, C03), CRC 1143 (project A05), and TRR 80 (project E06), respectively, and the research grant JE 748/1.
The numerical simulations were performed on the JUWELS cluster at FZ J\"ulich and the CHEOPS cluster at RRZK Cologne.
\end{acknowledgments}

\appendix

\section{Correlations in the extended Kitaev model}
\label{sec:appendix:pffrg}

In this appendix, we investigate the effect of spin interactions beyond the pure Kitaev model on the spin-spin correlations. We individually evaluate the effect of nearest neighbor and next-nearest neighbor Heisenberg interactions as well as off-diagonal $\Gamma_1'$ interactions, as defined in Eq.~\eqref{eq:appendix:pffrg:model}, which are arguably relevant in \nair{} \cite{Winter17rev}. 
We make use of a pseudo-fermion functional renormalization group (pf-FRG) approach \cite{Reuther2010}, which has previously been employed to study the Kitaev model on the honeycomb lattice augmented by nearest neighbor \cite{Reuther2011c} and further neighbor \cite{Singh2012} Heisenberg interactions. 
Recently, the method has been extended to comprise arbitrary non-diagonal two-spin interactions \cite{Buessen2019}, including also interactions of $\Gamma_1'$ type. 
The pf-FRG approach can be understood as a simultaneous large-S expansion \cite{Baez2017}, typically associated with magnetically ordered phases, and large-N expansion \cite{Buessen2018a,Roscher2018}, commonly favoring spin liquid ground states. 
The treatment of both channels on equal footing explains why the method can be readily employed to model the delicate interplay of magnetic order and spin-liquid physics, both in two-dimensional \cite{Reuther2010} and three-dimensional \cite{Iqbal2016,Buessen2016} frustrated quantum spin systems. 

\begin{figure*}
	\centering
	\includegraphics[width=\linewidth]{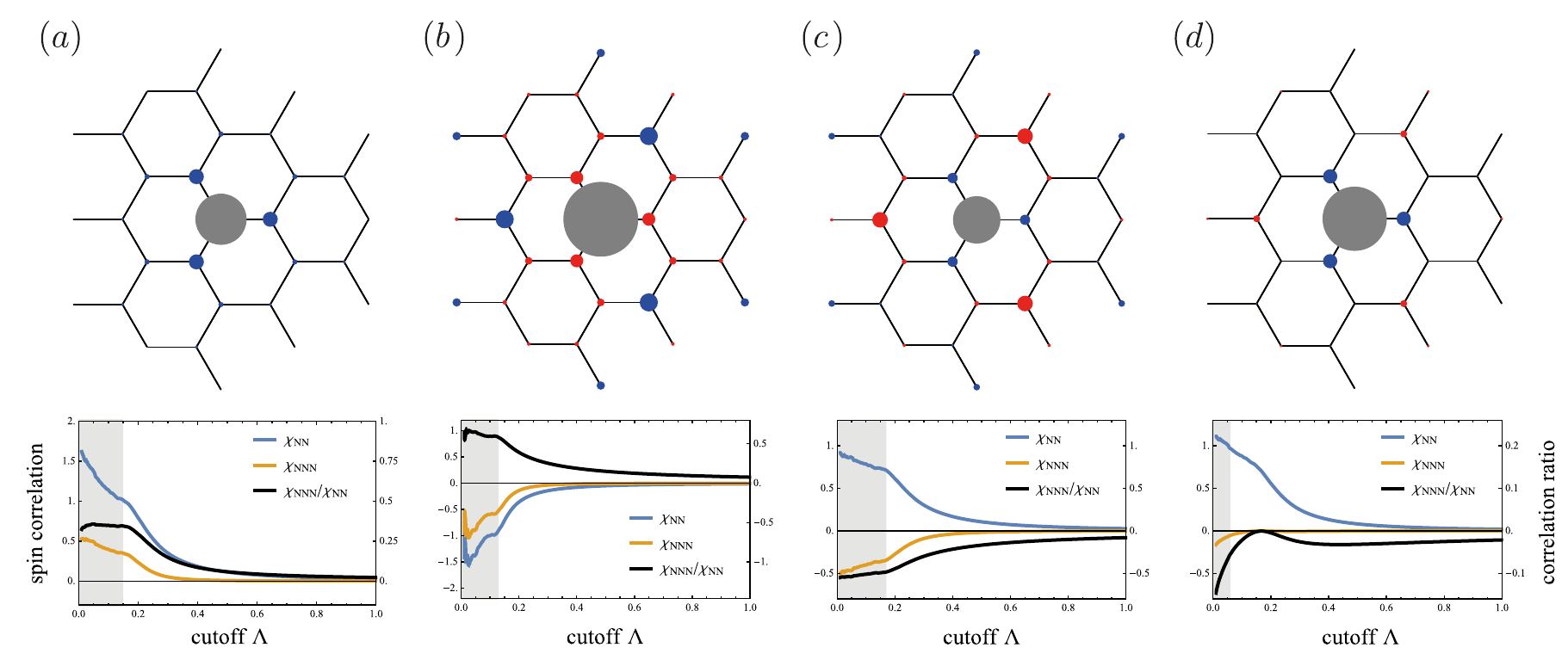}
	\caption{{\bf Correlations in the extended Kitaev model.} The upper row displays the real-space resolved spin-spin correlations just above the magnetic ordering transition. All correlations are measured with respect to the reference site indicated by the gray circle. The strength of the correlation is indicated by the size of the respective circle, while the coloring indicates the sign of the correlation; blue indicates ferromagnetic correlation and red is anti-ferromagnetic. 
		The bottom row shows the average nearest neighbor correlations $\chi_{NN}$ and next-nearest neighbor correlations $\chi_{NNN}$, as well as their ratio. The flow breakdown (see text for details) is indicated by the shaded region. 
		The panels are computed at different sets of coupling constants, all of which have additional ferromagnetic Kitaev coupling $K_1=-1$. 
		(a) $\Gamma'=-0.25$.
		(b) $J_1=0.5$.
		(c) $J_3=0.25$.
		(d) $J_1=0.1$, $J_3=0.05$, and $\Gamma_1'=-0.25$. }
	\label{fig:pffrgCorrelations}
\end{figure*}

The technique is based on a re-writing of spin operators in terms of Abrikosov fermions \cite{Abrikosov1965} (pseudo-fermions), and applying a conventional fermionic functional renormalization group (FRG) scheme \cite{Wetterich1993} to the resulting strongly coupled pseudo-fermion system. 
Within the FRG scheme, it is possible to simultaneously compute the renormalization (in a Wilsonian sense) of millions of interaction parameters. 
The renormalization group flow of the interaction parameters is generated by the installation of a (Matsubara-) frequency cutoff $\Lambda$; the flow equations are solved numerically from the limit of infinite cutoff, which corresponds to high temperatures \cite{Iqbal2016}, down to zero cutoff where the fully interacting physical system at zero temperature is restored. 
If the system undergoes a magnetic ordering transition at finite temperatures, the associated spontaneous symmetry breaking manifests itself in the form of a non-analyticity in the flow of the spin-spin correlation function
\begin{equation}
\chi_{ij}^\Lambda = \int\mathrm{d}t \langle \mathbf{S}_i(t)\mathbf{S}_j(0) \rangle 
\end{equation}
at some critical cutoff scale $\Lambda_c$.
Properties of the system below the flow breakdown point $\Lambda_c$ cannot be resolved within pf-FRG, but the spin-spin correlations just above the breakdown can be used as an indicator for the type of magnetic order that is about to proliferate. 
Note that this definition of the correlation function corresponds to the elastic $(\omega$\,=\,$0)$  
component of the (real-space transformed) dynamic structure factor. 

Our general model of interest comprises Kitaev interactions at strength $K_1$, nearest and third-nearest neighbor Heisenberg interactions of strength $J_1$ and $J_3$, respectively, as well as $\Gamma_1'$ interactions. It is governed by the Hamiltonian
\begin{align}
\label{eq:appendix:pffrg:model}
H &= K_1\sum\limits_{\langle i,j\rangle^\gamma_1} S_i^\gamma S_j^\gamma 
+ J_1\sum\limits_{\langle i,j\rangle_1} \mathbf{S}_i \mathbf{S}_j 
+ J_3\sum\limits_{\langle i,j\rangle_3} \mathbf{S}_i \mathbf{S}_j \nonumber\\
&\quad + \Gamma_1'\sum\limits_{\langle i,j\rangle_1^\gamma} \sum\limits_{\alpha \neq \gamma} \left( S^\gamma_i S^\alpha_j + S^\alpha_i S^\gamma_j \right) \,,
\end{align}
where $\langle i,j\rangle_1$ and $\langle i,j\rangle_3$ denote summations over all nearest neighbors and third-nearest neighbors, respectively, and the index $\gamma$\,=\,$x$, $y$, $z$ denotes the bond type. 
We investigate four different scenarios of spin interactions beyond the pure ferromagnetic Kitaev model ($K_1$\,=\,$-1$). 
(i)~Augmenting the Kitaev model with a finite $\Gamma_1'$\,=\,$-0.25$ generates ferromagnetic spin-spin correlations on nearest, next-nearest and further neighbors. It induces a phase transition into a ferromagnetically ordered state, indicated in our pf-FRG calculations by a kink in the RG flow of the observables, below which they exhibit unphysical behavior (Fig.~\ref{fig:pffrgCorrelations}a). 
(ii)~Adding sizable anti-ferromagnetic nearest neighbor Heisenberg interactions $J_1$\,=\,$0.5$ drives the system into the phase of `stripy' order. Its symmetrized correlations $\chi_{ij}^\Lambda$, which we determine just above the phase transition at $\Lambda_c$, are anti-ferromagnetic for nearest neighbors as well as next-nearest neighbors (Fig.~\ref{fig:pffrgCorrelations}b). 
(iii)~The combination of Kitaev interactions with anti-ferromagnetic third-nearest neighbor Heisenberg exchange $J_3=0.25$ favors a `zigzag' ordered ground state. This configuration has ferromagnetic nearest-neighbor correlations, while the second-nearest neighbors correlate anti-ferromagnetically (Fig.~\ref{fig:pffrgCorrelations}c). 
(iv)~{\em Ab-initio} calculations \cite{Yamaji14} suggest the presence of anti-ferromagnetic nearest-neighbor Heisenberg interactions $J_1/|K_1|\approx 0.1$ and third-nearest neighbor Heisenberg interactions $J_3/|K_1|\approx 0.05$ in \nair{}. We have argued that both of these couplings favor anti-ferromagnetic correlations on next-nearest neighbor level. These correlations can be suppressed in the presence of finite $\Gamma_1'$ interactions. At moderate values for $\Gamma_1'/|K_1| \approx -0.25$, the next-nearest neighbor correlations thereby almost vanish (Fig.~\ref{fig:pffrgCorrelations}d). 
At smaller values of $\Gamma_1'$ the suppression is less drastic, particularly at $\Gamma_1'/|K_1|\approx 0.1$ which is the value suggested by {\em ab-initio} calculations; in our calculations, however, we neglect bond anisotropies in the coupling constants that are also part of the {\em ab-initio} theory. 
Nevertheless, we have qualitatively demonstrated that the presence of $\Gamma_1'$ interactions in \nair{} may play an important role in the suppression of next-nearest neighbor correlations at temperatures above the ordering transition. 


\begin{thebibliography}{99}
\bibitem{FootnoteDisorder}
However, disorder related to the H ions may play an essential role in H$_3$LiIr$_2$O$_6$ 
\cite{Yadav18,Li18,Knolle19b}.
\end{thebibliography}
\end{document}